\documentclass[manuscript]{acmart}

\usepackage{lscape} 
\usepackage[normalem]{ulem}
\useunder{\uline}{\ul}{}
\usepackage{natbib}
\usepackage{amsmath}

\usepackage{amssymb}
\usepackage{graphicx}
\usepackage{subfigure}
\usepackage{caption}
\usepackage{rotating}

\AtBeginDocument{%
  \providecommand\BibTeX{{%
    \normalfont B\kern-0.5em{\scshape i\kern-0.25em b}\kern-0.8em\TeX}}}

\setcopyright{acmcopyright}
\copyrightyear{2018}
\acmYear{2018}
\acmDOI{10.1145/1122445.1122456}

\acmConference[Woodstock '18]{Woodstock '18: ACM Symposium on Neural
  Gaze Detection}{June 03--05, 2018}{Woodstock, NY}
\acmBooktitle{Woodstock '18: ACM Symposium on Neural Gaze Detection,
  June 03--05, 2018, Woodstock, NY}
\acmPrice{15.00}
\acmISBN{978-1-4503-XXXX-X/18/06}

\begin{document}

\title{Conversational Agents in Therapeutic Interventions for
Neurodevelopmental Disorders: A Survey}
\author{Fabio Catania}
\authornote{Both authors contributed equally to this research.}
\email{fabio.catania@polimi.it}
\orcid{0000-0002-5403-9002}
\author{Micol Spitale}
\authornotemark[1]
\email{micol.spitale@polimi.it}
\orcid{0000-0002-3418-1933}
\affiliation{%
  \institution{Politecnico di Milano}
  \streetaddress{Via Camillo Golgi 39}
  \city{Milan}
  \country{Italy}
}

\author{Franca Garzotto}
\orcid{0000-0003-4905-7166}
\affiliation{%
  \institution{Politecnico di Milano}
  \streetaddress{Via Camillo Golgi 39}
  \city{Milan}
  \country{Italy}}
\email{franca.garzotto@polimi.it}

\renewcommand{\shortauthors}{Catania and Spitale, et al.}


\begin{abstract}
    Neurodevelopmental Disorders (NDD) are a group of conditions with onset in the developmental period characterized by deficits in the cognitive and social areas. Conversational agents have been increasingly explored to support therapeutic interventions for people with NDD. This survey provides a structured view of the crucial design features of these systems, the types of therapeutic goals they address, and the empirical methods adopted for their evaluation. From this analysis, we elaborate a set of recommendations and highlight the gaps left unsolved in the state of the art, upon which we ground a research agenda on conversational agents for NDD.
\end{abstract}



\maketitle

\section{Introduction}
\label{introduction}
\textit{Neurodevelopmental Disorders} (NDD) are a group of conditions that originate during childhood and are characterized by severe deficits in the cognitive and social areas \cite{american2013diagnostic}. NDD is a chronic state, but early and focused interventions are thought to mitigate its effects \cite{dawson2012early}.
Traditional therapy involves different approaches for assessing  NDD symptoms and helping persons with NDD develop solid social skills \cite{astramovich2015play, kim2009emotional}, learn receptive and expressive language \cite{flippin2010effectiveness}, improve adaptive behavior, and promote autonomy in self-care and everyday life tasks. 

\textit{Conversational technology} is defined as any software that can provide access to information and services through natural language \cite{folstad2020users, hobert2019say} and has recently been identified as a potentially beneficial means to complement more traditional interventions for people with NDD since the interaction is perceived as predictable, "safer" and more straightforward compared to communication with other humans \cite{huskens2013promoting, rafayet2018virtual, tanaka2017embodied, allen2018echo, dickerson2013action, chevalier2017dialogue}. 

Given the broad spectrum of special needs to be addressed in people with NDD and the very multidisciplinary aspects to consider when designing and evaluating conversational agents for NDD, it is not easy for interaction designers and researchers to have a clear picture of their benefits or drawbacks and the open questions to address.
In this context, we asked: \textit{"How have conversational agents been used to support the therapy of people with NDD, and to what extent are the chosen design features and methods suitable to assess and train their skills?"}.
To provide a multi-faceted answer of interest to both designers and researchers, we asked then the following sub-questions:

\begin{enumerate}
\item[R1] How many participants are involved on average in the studies with conversational agents and people with NDD? 
\item[R2] What is the age of the participants? 
\item[R3] What is their gender? Are males and females equally included?
\item[R4] What is their diagnosis?
\item[R5] Which participant's skills are addressed within these studies?
\item[R6] Are conversational agents used for assessing or for training the skills of people with NDD?
\item[R7] What interaction modalities are exploited as input and output by conversational agents for NDD?
\item[R8] What "wake actions" (i.e., operations by the user to signify the beginning of each interaction) are used by these conversational agents? 
\item[R9] What is the embodiment of these agents (i.e., their representation or expression in a tangible or visible form)?
\item[R10] What is their shape?
\item[R11] What is the gender of these conversational agents?
\item[R12] Are the conversational agents endowed with sentiment or emotion recognition capabilities?
\item[R13] 
Are conversational agents' prototypes automated, semi-automated, or manually controlled in the studies for NDD?
\item[R14] Are the therapy sessions with conversational agents task-based or participants with NDD can freely interact with the technology? 
\item[R15] Which study designs are used in the studies with conversational agents and people with NDD?
\item[R16] How long do these studies last?
\item[R17] What data collection methods are used?
\item[R18] What are the main lessons learned in terms of empirical research method in the studies with conversational agents and people with NDD?
\item[R19] What are the main lessons learned in terms of conversational agents' design highlighted by the current state of the art?
\item[R20] What are the main lessons learned in terms of technology?
\item[R21] 
What are the main research questions left unsolved in the field? 
\end{enumerate}

To address all these questions, we performed a literature review, and we did it systematically to minimize the effect of data collection, selection, and extraction bias.

The major contributions of this work are threefold:
\begin{itemize}
\item we provide a systematic synthesis of the state of the art on conversational agents for NDD therapy organized along different dimensions (i.e., users' profile, therapeutic goals of the conversational agents, design features of the agents, methods and procedures used in the empirical evaluation studies, studies' results and unresolved research questions);
\item by distilling the lessons learned from the studies in the literature review, we suggest some recommendations for interaction designers and researchers about key features to take into account during the design process and the empirical evaluation of conversational agents for the therapy of people with NDD. 
Particularly, we propose a checklist for HCI researchers who want to run an empirical study with a conversational agent for NDD to ensure a high-quality report in their papers without missing the most relevant information about the design features of the agent and the methodology aspects of the study;
\item we draw a research agenda that encompasses  the directions for future work highlighted in the various studies and helps researchers in the field of conversational agents for NDD to identify priorities and challenges to address in their future work.
\end{itemize}

The rest of the paper is organized as follows. 
In Section \ref{ndd}, we provided some background material about the neurodevelopmental disorders and their wide variability.
Section \ref{methodology} overviews the choices we took and the procedural steps we followed during the study.
Section \ref{results} describes the corpus of the papers selected for analysis and addresses each of the research questions mentioned above. 
Section \ref{discussion} present some design and methodology recommendations we elaborated from the lessons we learned from previous works and draws some trajectories for future research.
We end with Section \ref{conclusion} that sums up the results of this work.

\section{Neurodevelopmental disorder}
\label{ndd}

According to the fifth edition of the \textit{Diagnostic and Statistical Manual of Mental Disorders} (DSM-5) \cite{american2013diagnostic}, NDD is a group of conditions that originates in the developmental period and persists into adolescence and adulthood. It is characterized by different developmental deficits and produce impairments of personal, social, academic, or occupational functioning. The causes of NDD should be searched mainly in a combination of genetic and environmental factors, but much research is still ongoing in
this area \cite{martens2007genetic}. 
The umbrella term \textit{NDD} comprises \textit{intellectual developmental disorder}, \textit{global developmental delay}, \textit{communication disorders}, \textit{autism spectrum disorder} (ASD), \textit{attention-deficit/hyperactivity disorder} (ADHD), \textit{neurodevelopmental motor disorders}, and \textit{specific learning disorder}.

The boundaries between the different disorders are not strict, and comorbidities are very common. Frequently, one disorder overlaps and overshadows the traits of some others, making the diagnostic process more complex \cite{astle2021annual}. The neurodevelopmental continuum highlights the need for flexible approaches to diagnosis. For this reason, assessment is usually performed by merging different complementary approaches and considering multiple perspectives. It can be based on (i) direct observations of individuals' behavior in social contexts, (ii) analysis of their developmental history, and (iii) interviews with their caregivers \cite{steiner2012practitioner}. 

NDD prevalence estimations typically vary across reported sources, and comparisons are extremely hard \cite{cdc}. 
To provide some numbers, in 2015, approximately 15\% of people in the United States aged from 3 to 17 had a neurodevelopmental disorder diagnosis, with the highest prevalence of ADHD and learning disabilities \cite{Office_of_Research2009-da}. Moreover, some studies suggest that many disorders are still under-diagnosed in the population aged less than 18 years \cite{marino2018prevalence}, and so numbers are likely to rise over the years \cite{durkin2019increasing}.

Speaking about treatments, any neurodevelopmental disorder is chronic, but early and focused interventions are thought to mitigate its effect.
The neurodevelopmental continuum suggests that therapeutic approaches may be fruitful across diagnostic boundaries. Still, the ideal treatment includes regular interventions that meet a person's specific needs and objectives and monitor their development \cite{reichow2013non}. 
Objectives encompass strengthening adaptive behaviors, developing association skills and abstract concepts (e.g., time and money), expanding emotional and relational skills \cite{astramovich2015play, kim2009emotional}, learning receptive and expressive language \cite{flippin2010effectiveness}, enhancing communicative and speech skills, and improving self-care and autonomy. 
From literature we know that Standard therapeutic interventions for people with NDD include different approaches such as play therapy \cite{astramovich2015play, hillman2018child}, music therapy \cite{warwick1991music}, speech therapy \cite{hoque2009exploring}, occupational therapy \cite{watling1999current}, Cognitive Behavior Therapy (CBT) \cite{hronis2017review, danial2013cognitive}, and practices based on Applied Behavior Analysis (ABA) \cite{axelrod2012applied, foxx2008applied}.

Nowadays, high costs preclude many people with NDD from accessing proper treatment \cite{Oono2013-sv}. In addition, NDD often imposes a significant emotional, psychological, and organizational burden on people with these disorders and their families. Caring for a person with a severe form of NDD may be demanding, especially where access to services and support are inadequate \cite{who, hansen2015explaining}. For example, the cost of supporting an individual with an ASD and intellectual disability throughout the lifespan is \$2.4 million in the United States and \$2.2 million in the United Kingdom. The cost of supporting an individual with an ASD without intellectual disability is \$1.4 million in the United States and \$1.4 million in the United Kingdom \cite{buescher2014costs}.
In this context, integrating traditional interventions with interactive technologies, in general, and conversational technology, in particular, can increase the benefits of therapy and make it more accessible, frequent, and affordable.

\section{Method}
\label{methodology}
We systematically reviewed the literature to address the research questions stated in Section \ref{introduction} and obtain a comprehensive analysis and meta-analysis of how conversational agents had been used in past works for the therapy of people with NDD and to what extent the approaches taken were suitable to assess and train their skills. 
In light of the complexity of our research challenge, we selected a number of experimental studies which are relevant in this domain, and we extracted and analyzed mixed forms of data, including both quantitative and qualitative findings.

In Section \ref{defterms}, we defined some terms about conversational technology that we considered relevant for our survey. In this way, we wanted to improve the readability of the paper and prevent language-related misunderstandings in a domain where concepts are sometimes defined in different ways depending on the context. Then, in Section \ref{procedure}, we detailed our papers' selection and analysis process.

\subsection{Defining terms}
\label{defterms}

\subsubsection{Conversational agents}
\textit{Conversational agents} are software providing access to information and services through natural language (e.g., English), in line with the definitions of \citet{folstad2020users} and \citet{hobert2019say}. 
Sometimes \textit{dialog system} and \textit{chatbot} are used as synonyms of conversational agent \cite{conversation2021}.
In our survey, we focused on a particular category of conversational agents characterized by the capability of managing speech-based user's input.
Therefore, in this paper we use the term "conversational agent" only to refer to this category of systems. 

\subsubsection{Embodiment}
The \textit{embodiment} of a conversational agent is its representation or expression in a tangible or visible form \cite{shape}.
Previous work on conversational agents is situated in long standing fields of research exploring different types of embodiment.
In the context of this study, we considered disembodied conversational agents or DCAs (i.e., agents with no dynamic physical representation \cite{araujo2018living}), embodied conversational agents or ECAs (i.e., virtual agents on tablet or computer screens with abstract, cartoon-like or human-like appearance \cite{cassell2001embodied}), socially assistive robots or SARs (i.e., physical agents for social interaction \cite{feil2005defining}), and Intelligent Personal Assistants or IPAs (i.e., smart assistants allowing natural language interaction \cite{garrido2010adding}).

\subsubsection{Shape of the agent}
There are some theories classifying different social agents with respect to the shape of their embodiment. For example, according to \citet{fong2003survey} and \citet{shibata2004overview}, the shape of a socially assistive robot can be defined as \textit{bio-inspired} when it simulates the social intelligence found in living creatures (including humanoids, animals, and vegetables), as \textit{functional} when it reflects the task it performs, or as \textit{artificial} when it shows artificial (e.g., mechanistic) characteristics.
For the scope of this review, we considered the above-mentioned definitions suitable for conversational agents in general and opted for using them in this paper. 

\subsubsection{Wake action}

Normally, before the interaction with the conversational agent begins and whenever the agent ends a conversational turn, it lies in an idle state and can only sense a specific wake action. As soon as it recognizes that action, the agent gets triggered and listens to the user's speech. Next, it stops processing the input when it recognizes a pause.
So, we define a \textit{wake action} as an operation performed by the user to signify the beginning of interaction with the agent or to express the concept "now it is my turn" during the conversation.
Wake actions are believed to promote the sense of agency and increase user's subjective awareness of being in control of the interaction \cite{agency}.
Wake actions can be of various nature. For example, they can be:
\begin{itemize}
    \item \textit{vocal}, when the user utters a phrase that prompts the device to begin processing the speech. For example, popular conversational assistants like Amazon Alexa, Google Assistant, and Apple Siri are activated by pronouncing their wake word ("Alexa", "OK Google", and "Hey Siri", respectively), and they stop listening when they recognize a pause;
    \item \textit{tactile}, when the user presses a button (physical or digital) or toggles a control to trigger the system;
    \item \textit{visual}, when the user uses the gaze to communicate the intention to speak;
    \item \textit{event-based}, when pre-determined events (e.g., an alarm) wake up the system;
   \item \textit{motion-based}, when the user prompts the system by moving parts of the body (e.g., waving the hand in front of a sensor).
\end{itemize}

\subsubsection{Sentiment and emotional analyses}
Conversational agents may be empowered with some typically human abilities of interpreting and expressing "feelings" so as to adapt the interaction and make it appear more emphatic and natural.
For example, adding \textit{sentiment analysis} capabilities enable the agent to find out whether a given input expresses positivity, neutrality, or negativity, and to take into account this information to generate a more personalized output. 
Usually, sentiment analysis outputs a number between 0 (totally negative) and 1 (totally positive). Differently, \textit{emotional analysis} consists in eliciting the user's emotions (e.g., joy...) during the conversations by processing linguistic or audio features of their input.

\subsubsection{Automated, semi-automated, and controlled prototypes}
Conversational agents are \textit{automated} if they work autonomously, \textit{semi-automated} if they are generally autonomous but require the real-time control by a human for some specific scenarios and functionalities, and \textit{controlled} if they totally depends on the decisions and actions of a human being.
For the evaluation of conversational agents in the field of NDD, previous works have used a wide variety of approaches, including automated, semi-automated, and
\textit{"Wizard-of-Oz" prototypes} \cite{frauenberger2010phenomenology, hoysniemi2004wizard}. The latter ones simulate autonomous behaviours but are just mock-ups manually controlled by a researcher or designer.

\subsection{Procedure}
\label{procedure}
To outline our systematic procedure for reviewing the literature, we followed the guidelines by \citet{nightingale2009guide} and got inspired by another review on conversational agents for a generic audience \cite{seaborn2021voice}.

\begin{figure}
    \centering
    \includegraphics[width=0.6\textwidth]{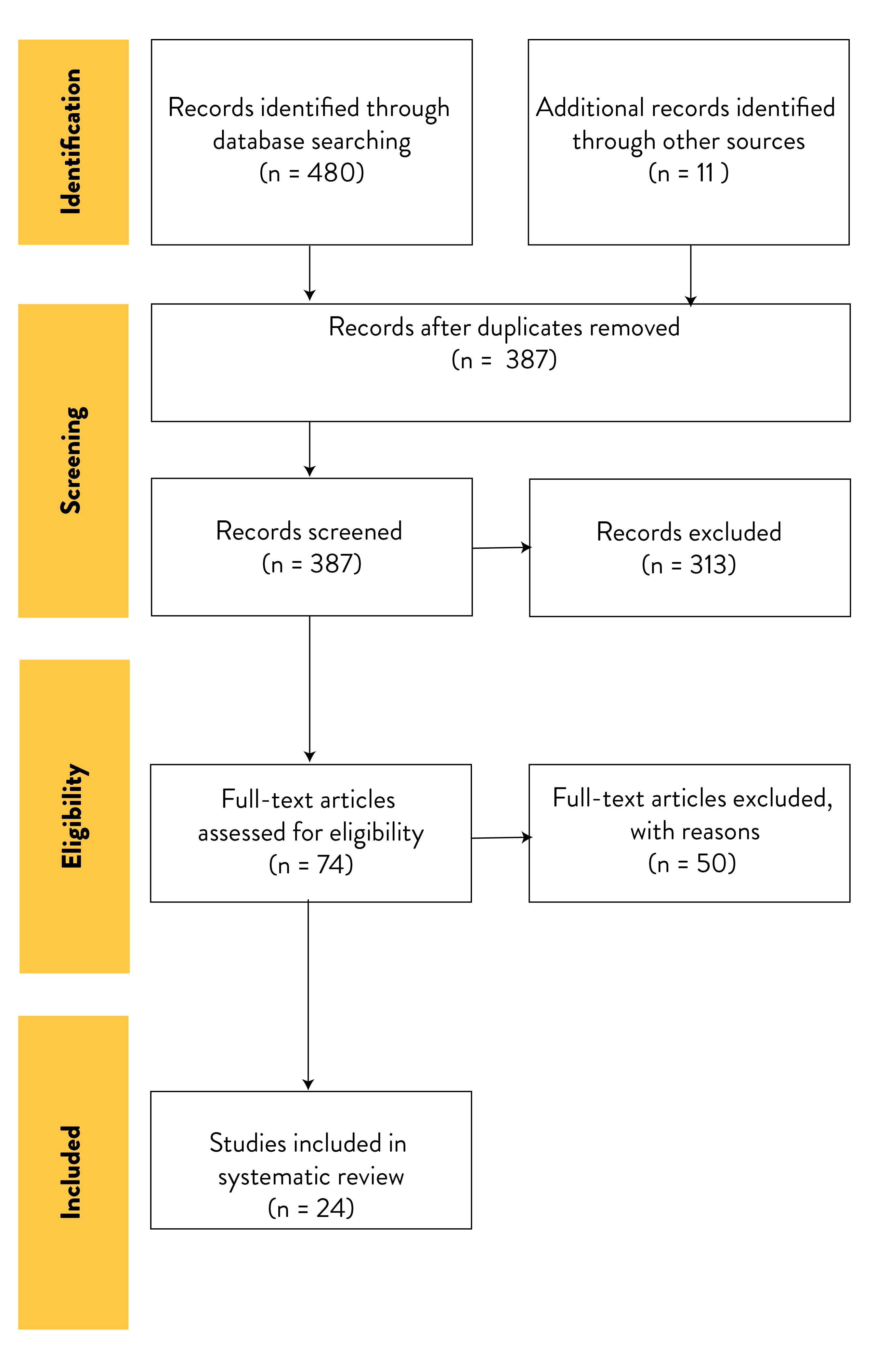}
    \caption{Flow diagram of our systematic review in the PRISMA format.}
    \label{fig:prisma}
\end{figure}

Our workflow followed the PRISMA format \cite{moher2009preferred}. PRISMA stands for Preferred Reporting Items for Systematic Reviews and Meta-Analyses and is recognized as the state-of-the-art method for systematic reviews and meta-analyses since proves the quality of the review, permits its replication, structures the final manuscript using standard headings, and allows readers to assess strengths and weaknesses of the study \cite{nightingale2009guide}.
More in detail, the PRISMA format provides an evidence-based minimum set of items for reporting in systematic reviews and consists of the 4-phase flow shown in Figure \ref{fig:prisma}. 
The first stage is \textit{identification} and describes when all candidate manuscripts are collected.
Then, in the \textit{screening} phase, only the papers that match the eligibility criteria are filtered in based on the analysis of their titles and abstracts.
During the subsequent \textit{selection} phase, full texts are examined and only those papers that still meet the same eligibility criteria as above are included.
Finally, during the \textit{inclusion} phase, all the papers that passed all the selection steps are analyzed to address the research questions of the study.

Following, we present our eligibility criteria for the review, the details of our selection process, our search query keywords, and the methods used for data extraction and analysis.

\subsubsection{Eligibility criteria}
We developed a set of inclusion and exclusion criteria for our literature review based on the research questions and the guidelines by \citet{kitchenham2004procedures} for selecting papers in engineering. 

Papers were included if:
\begin{itemize}
    \item they involve people with NDD,
    \item they address the assessment or training of any social, communication, language, or emotion skills, 
	\item they cover the description of an empirical study,
	\item they employ a conversational agent,
	\item their title, abstract, and keywords contain at least one keyword describing such technology and one keyword defining the target population from the Search Query Keywords (see Section \ref{searchquerykeywords}).
\end{itemize}

Papers were excluded if:
\begin{itemize}
    \item they were published before 2011 and after the day of actual running of the research, i.e., June 7, 2021,
	\item they are not in English,
	\item they are not in peer-reviewed journals and conference proceedings,
	\item they are survey papers,
	\item they are inaccessible to the authors,
	\item they are low quality (i.e., they do not report the details necessary to evaluate their eligibility).
\end{itemize}

Also, if multiple papers describe the same empirical study, only one of the papers was kept for further analysis.

\subsubsection{Selection process}
Papers were first screened at a high level and then selected after in-depth analysis.
Two independent reviewers screened the titles and abstracts on the eligibility criteria mentioned above. 
We assessed the inter-rater reliability to check whether the two reviewers worked consistently. To do so, we calculated the Cohen’s Kappa statistic \cite{cohen1960coefficient}. 
When the two reviewers disagreed, a third reviewer was consulted and determined the decisive inclusion or exclusion of the papers.
Finally, the full text of the remaining papers was analyzed. The selection process was analogous to that described above. The same two reviewers of the screening phase assessed the full text versions of the remaining articles, and the inter-rater reliability was assessed again as explained above. The same third reviewer was consulted in case of disagreement and rated the final inclusion of the papers.

\subsubsection{Search query}
\label{searchquerykeywords}
Papers were collected based on the terms appearing in their title, abstract, and keywords. The list of terms we searched for described both the conversational technology and the target population. 
Terms about the technology included those we observed to be most common in the literature and are: \textit{conversational technology}, \textit{dialog system}, \textit{chatbot}, \textit{conversational agent}, \textit{embodied conversational agent}, \textit{virtual agent}, \textit{socially assistive robot}, and \textit{intelligent personal assistant}.
Terms about the target population refer to the neurodevelopmental disorders included in the DSM-5 \cite{american2013diagnostic}. They are: \textit{intellectual disabilities}, \textit{communication disorders}, \textit{autism spectrum disorder}, \textit{attention-deficit/hyperactivity disorder}, \textit{specific learning disorder}, \textit{motor disorders}, and the umbrella expression \textit{neurodevelopmental disorders}. Alternative term forms, such as pluralization, were considered.

We searched Scopus, ACM digital library, and IEEE explore databases.
Scopus was chosen given that it provides access to several databases with a wider coverage beyond the computer science field.
ACM digital library and IEEE Explore were selected because they capture respectively a substantial number of publications in the human-computer interaction and computer science and engineering fields. 
The queries differed slightly based on each database’s requirements. As an example, we report the search query on Scopus:

\begin{verbatim}
TITLE-ABS-KEY ( ( "conversational technolog*" OR "dialog* system*" OR "chatbot*" 
OR "conversational agent*" OR "embodied conversational agent*" OR "virtual agent*"
OR "social* robot*" OR "intelligent personal assistant*" ) 

AND

( "neurodevelopmental disorder*" OR "intellectual disabilit*" OR "communication disorder*"
OR "autism spectrum disorder*" OR "attention-deficit/hyperactivity disorder*" 
OR "specific learning disorder*" OR "motor disorder*" ) ).
\end{verbatim}

After collecting the possible papers to review by running the search queries, we manually added a number of potentially relevant papers that were not listed in the query results based on our previous knowledge of the domain. Next, we filtered out the duplicates and stored references in a Microsoft Excel file.

\begin{table}[h!]
\small
\centering
\caption{Research questions associated with variables' name, type, and approach exploited for the data's extrapolation.}
\label{tab:survey-rqs}
\begin{tabular}{clll}
\toprule
\textbf{Research} & \textbf{Variables}                                       & \textbf{Type} & \textbf{Approach}\\
\textbf{ Questions}&& \\
\midrule
R1                                             & Participants' number                                      & Numeric &  Analytic   \\
R2                                             & Participants' age                                        & Numeric   &   Analytic   \\
R3                                             & Participants’ gender (i.e., male, female)                                    & Categorical & Top-down   \\
R4                                             & Participants’ diagnosis (i.e., ID, ASD, etc. as in \cite{american2013diagnostic})                                  & Categorical &  Top-down   \\

R5                                             & Participants' skills addressed & Categorical  & Bottom-up  \\
                                            & (e.g., emotional, joint attention)  &  \\
R6                                             & Agent's goal  (i.e., train, assess)                                             & Categorical  & Top-down   \\
R7                                             & Input and output interaction modalities              & Categorical &   Bottom-up \\
                                            & (e.g., speech, gesture)  &  \\
R8                                            & Agent’s wake action                                      & Categorical  &  Bottom-up  \\
                                        & (e.g., buzzer, wake word, none)  &  \\
R9                                             & Agent’s embodiment        & Categorical  &  Top-down  \\
&(i.e., ECA, IPA, SAR, or disembodied)  &\\
R10                                            & Agent’s shape                                            & Categorical  &  Bottom-up  \\
&(i.e., humanoid, animal-like, speaker, vegetable )  &\\
R11                                             & Agent’s gender                                           & Categorical &   Top-down  \\
                                        & (i.e., male, female, neutral)  &  \\

R12                                            & Agent’s emotional recognition   capabilities             & Categorical  &  Bottom-up  \\
                                        & (e.g., facial expression)  &  \\
R13                                            & Nature of the prototype       & Categorical &   Top-down  \\
&(i.e., autonomous, semi-autonomous, or Wizard-of-Oz) &\\
R14                                          & Intervention Type    & Categorical   & Top-down  \\
&(i.e., task-based and free interaction) &\\
R15                                            & Study design       & Categorical  & Bottom-up  \\
&(e.g.,within-subject,   between-subject)&\\
R16                                            & Study duration                                           & Numerical   & Analytic   \\
R17                                           & Method of data collection  & Categorical &   Bottom-up \\
&(e.g.,interview, questionnaires)&\\

R18                                            & Study results about the empirical research method                           & Qualitative  & Pattern-based  \\
R19                                            & Study results about the design                           & Qualitative  &  Pattern-based \\
R20                                           & Study results about the technology                       & Qualitative &  Pattern-based  \\

R21                                            & Open research questions                                           & Qualitative &  Pattern-based \\
\bottomrule
\end{tabular}
\end{table}

\subsubsection{Data extraction and analysis}
We identified a variable for each of the twenty-one research questions. Variables are of different types: numerical (e.g., number of participants in each study), categorical (e.g., gender of the agent), or qualitative (e.g., study finding and lesson learned).
From every paper, we extracted information relevant to determine the value of each variable exploiting an \textit{analytic} approach (i.e., computing descriptive statistics) for numerical variables and identifying \textit{pattern-based} methods for the qualitative variables. The categorical variables have been obtained using either a \textit{top-down} or \textit{bottom-up} approaches. 
The former approach consists of defining the variable categories' a priori grounding them on the current state-of-the-art. For example, for R9, we checked the literature \cite{fong2003survey, shibata2004overview} and we chose a priori the possible categories of agent's shape to consider in the survey (i.e., bio-inspired, functional, and artificial). 
The latter approach refers to the extrapolation of the variable from the data collected among the selected papers. For example, for the R7 we obtained the different input and output modalities used in the surveyed studies and then we grouped them into categories (e.g., speech, gesture). 
The list of the variables and the corresponding research questions are shown in the Table \ref{tab:survey-rqs}.

\section{Totals and results}
\label{results}
The numbers of our paper selection are reported in Figure \ref{fig:prisma}.
A total of 480 papers were obtained by running the search query (387 after removing duplicates). Other 11 papers were manually added by the authors based on our previous knowledge.
After screening keywords, titles, and abstracts, reviewer 1 and reviewer 2 achieved a Cohen’s kappa coefficient of 81\%, indicating strong agreement between them \cite{cohen1960coefficient}. 74 papers were selected.
After reviewing full-text papers, the two reviewers achieved a Cohen’s Kappa rating of 89\%, indicating strong agreement again \cite{cohen1960coefficient}.

\textit{24 papers were finally included in the systematic review}. They are listed in Table \ref{tab:agent} and Table \ref{tab:agent2}, Table \ref{tab:interaction}, Table \ref{tab:demogr}, Table \ref{tab:skills}, Table \ref{tab:study}, and Table \ref{tab:data} of the Appendix \ref{appendix}.

As we can see from Figure \ref{fig:papartime}, the trend of the papers addressing conversational agents for the therapy of people with NDD is growing over years.

In this section, we report results grouped by sub-question. We only report descriptive statistics (frequency analysis) because our sample (24 papers) was to small to get any significance from statistical tests. 

\begin{figure}
    \centering
    \includegraphics[width=0.5\textwidth]{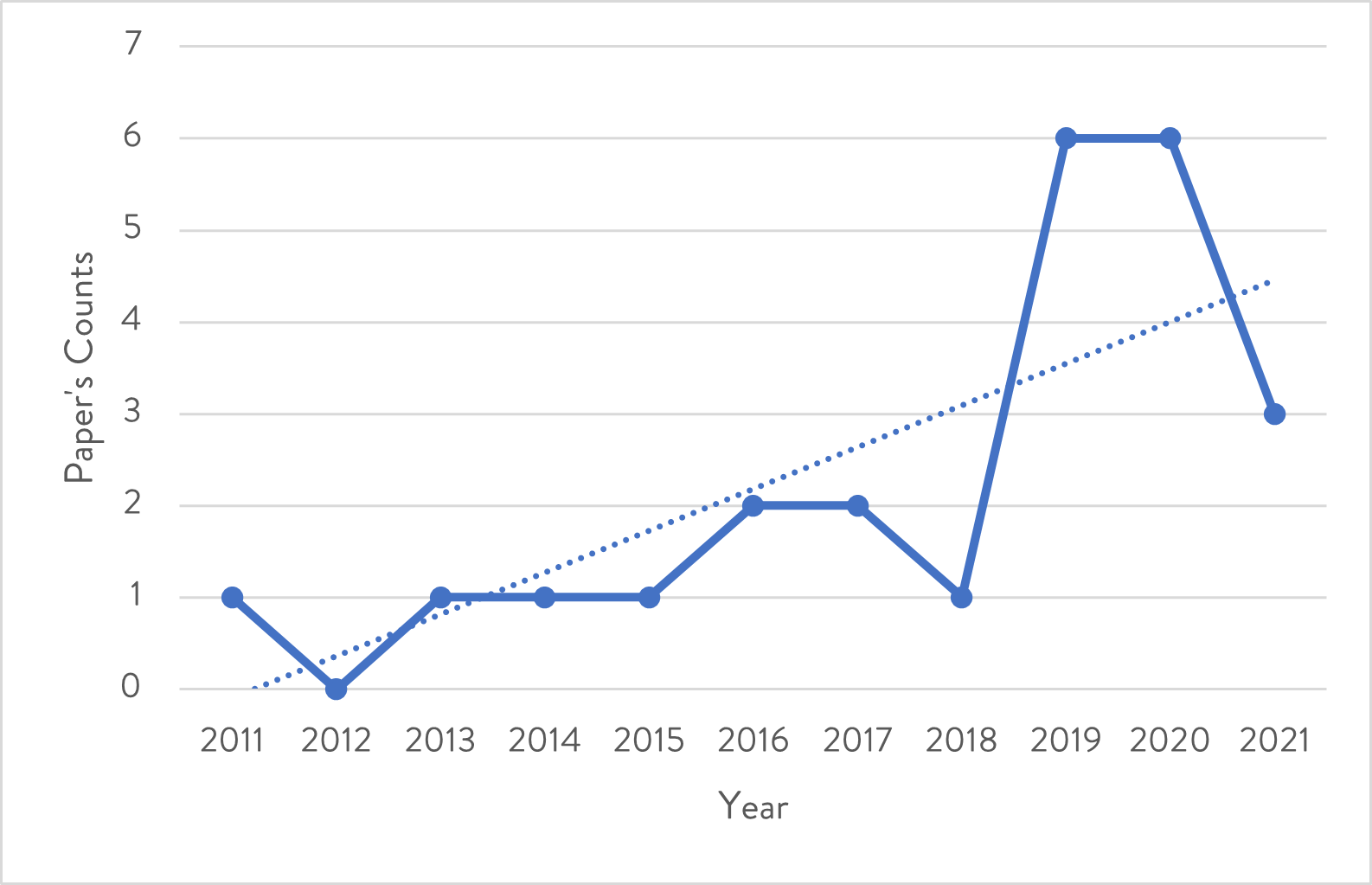}
    \caption{Number of papers over the last decade (from January 2011 to June 2021). The dot line represents the trend of the incremental interest in this field.}
    \label{fig:papartime}
\end{figure}

\subsection{Target group (R1, R2, R3, R4)}
\begin{figure}
\centering
\subfigure{\includegraphics[width=0.32\textwidth]{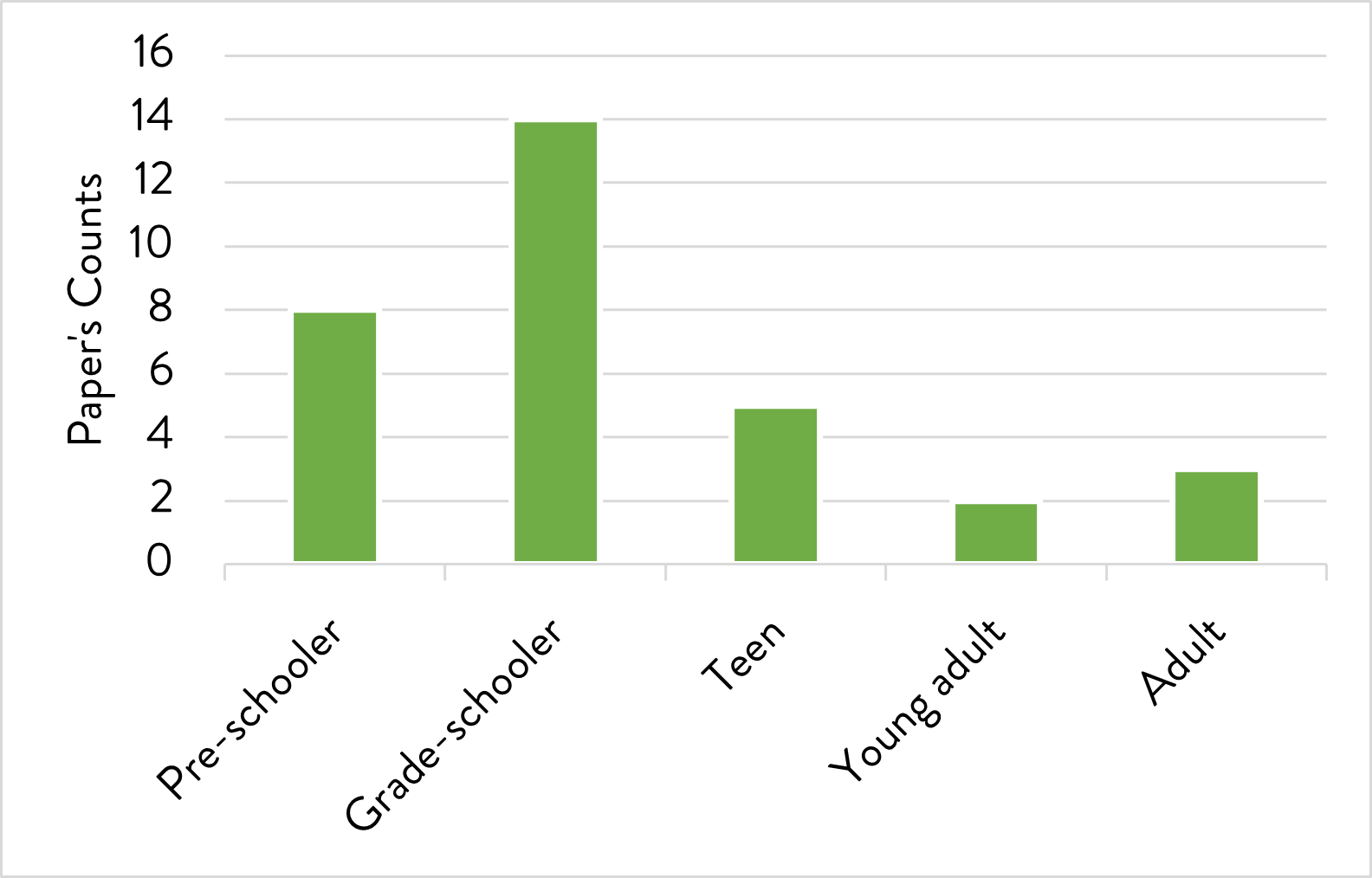}}
\subfigure{\includegraphics[width=0.32\textwidth]{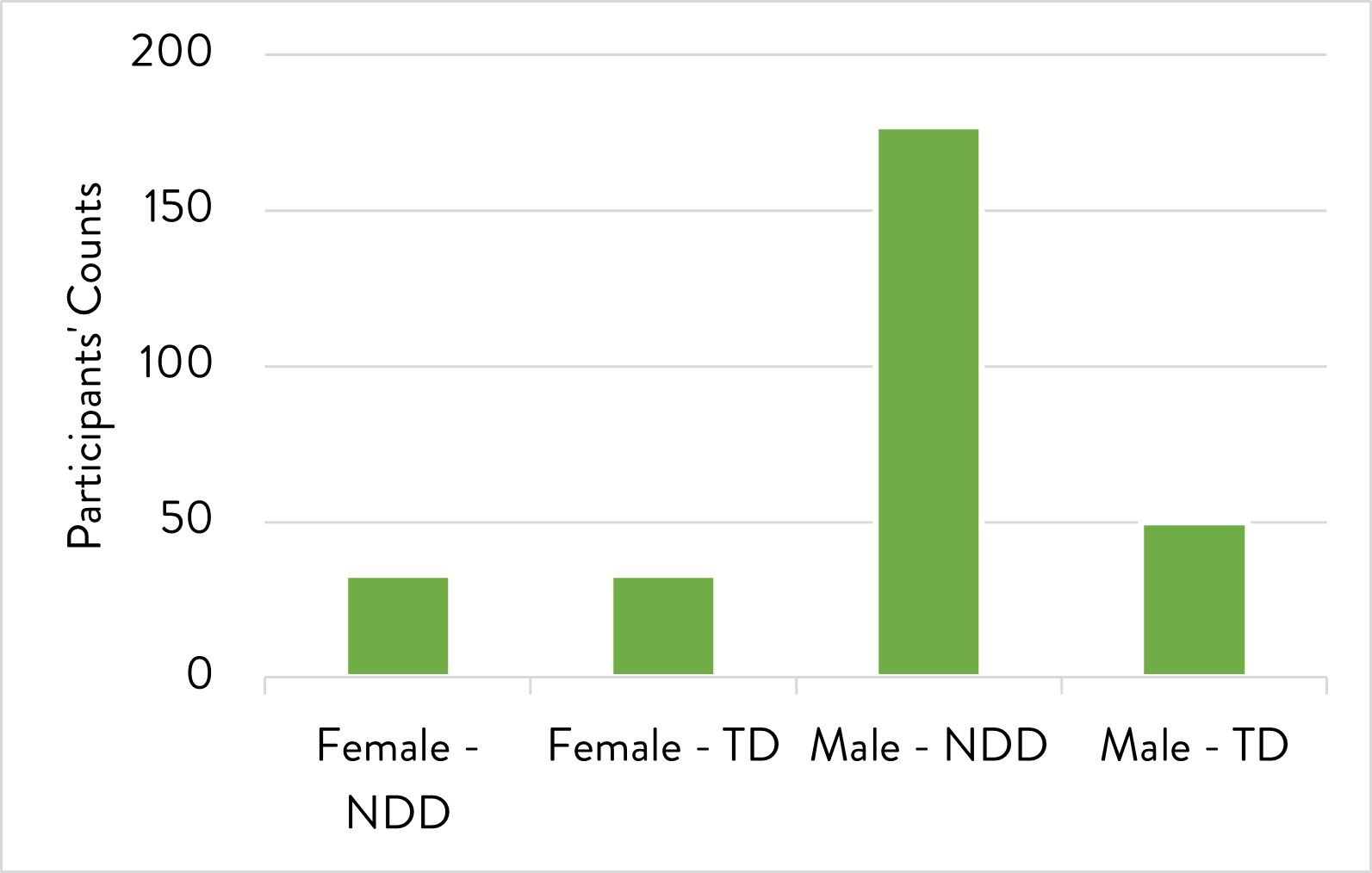}}
\subfigure{\includegraphics[width=0.32\textwidth]{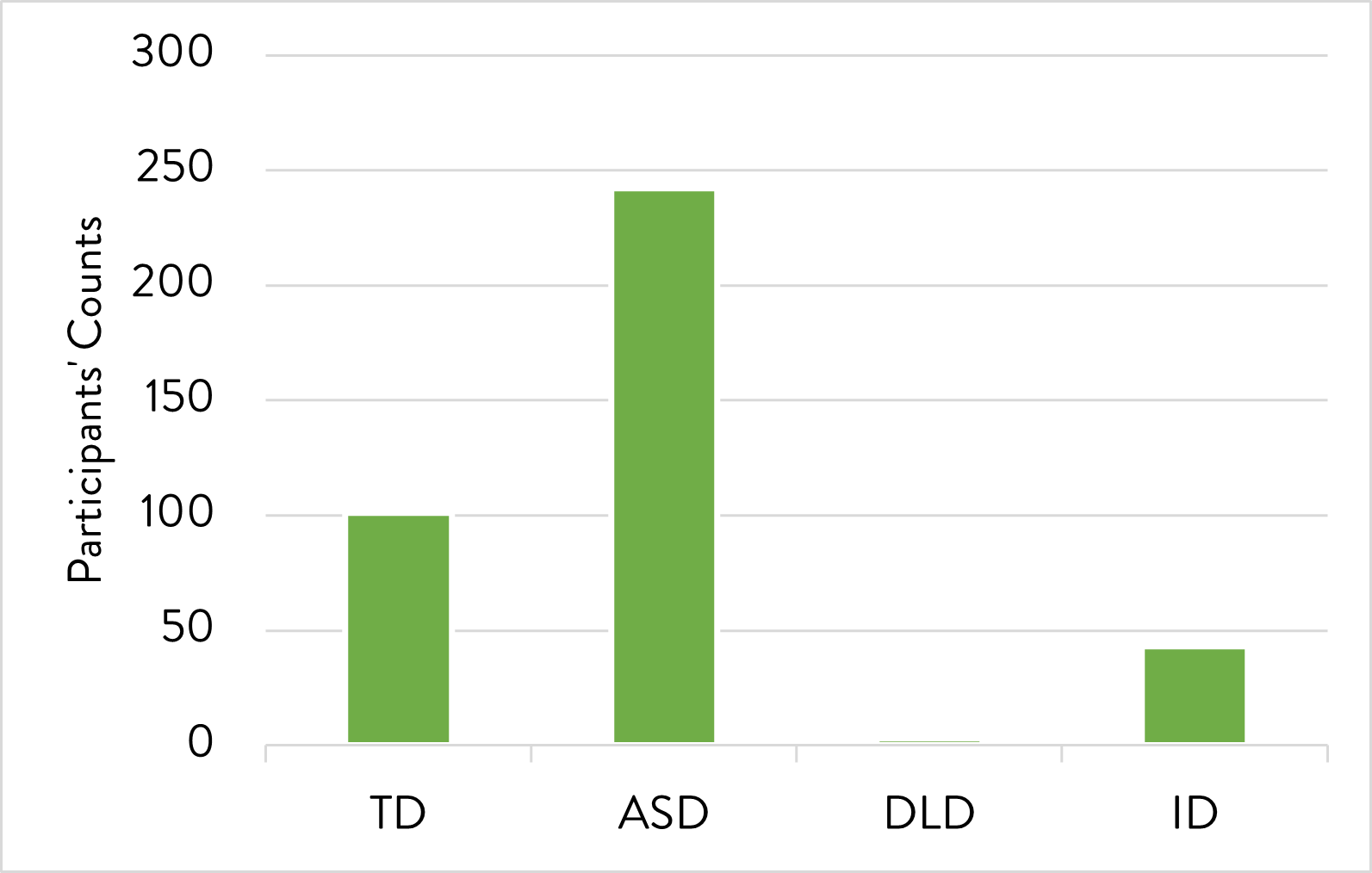}}
\caption{Target Group (R1, R2, R3, R4): participant's age (left), participant's gender (center), and participant's diagnosis (right).\\ \textit{Legend}: typically-developed (TD), neurodevelopmental disorder (NDD), autism spectrum disorder (ASD), intellectual disability (ID), and developmental language disorder (DLD). }
\label{fig:targetgroup}
\end{figure}


A total number of 367 participants were involved in the 24 studies included in this survey. 266 (72\%) participants were people with NDD while the remaining 101 (26\%) were typical developed (TD) people. The average number of participants with NDD for each study was 11.08 (SD = 9.58), while the TD participants were on average 16.83 (SD = 8.87) (see Figure \ref{fig:targetgroup}).
Regarding the specific diagnosis of the participants with NDD, 22 works (92\%) explored the use of conversational agents for people with autism spectrum disorders. Only one of those studies (4\%) included participants with autism with an associated diagnosis of intellectual disability. Only one study (4\%) included people with a primary diagnosis of intellectual disabilities, and another one (4\%) involved children with developmental language disorders (see Figure \ref{fig:targetgroup}).

About the demographics of the population involved in the surveyed studies, one paper (4\%) did not provide any information about the participants involved neither in terms of age or gender. Other four studies (16\%) only provided information about age but not gender.
21 studies (87\%) investigated the use of conversational agents for children (aged less than 18 years old). In particular, 8 studies involved pre-schooler children (3-6 years old), 14 of them included grade-schooler (6-12 years old), and 5 of them teens (12-18 years old), as shown in Figure \ref{fig:targetgroup}.
2 studies involved young adults (18-21 years old), and only 3 studies involved adults (more than 21 years old).
Studies generally included a specific target population age. 
Only 6 studies (25\%) included participants with a very broad range of age (equal or greater than 12 years range), that means for example children from pre-schooler to teens, and consequently a wide range of needs.
Considering all the studies of this survey, 177 males with NDD (48\%), 33 females with NDD (9\%), 50 TD males (14\%) and 33 TD females (9\%) were included. The gender of 74 participants (20\%) was not specified in the papers, among which 72 (19.6\%) with NDD and 2 (0.4\%) TD (see Figure \ref{fig:targetgroup}).

To sum up, \textit{results showed that researchers have mostly focused on male children (aged less than 18 years old) with ASD}.

\subsection{Participant's skills addressed and agent's goal (R5, R6)}
\label{goalandskills}

\begin{figure}
\centering
\subfigure{\includegraphics[width=0.32\textwidth]{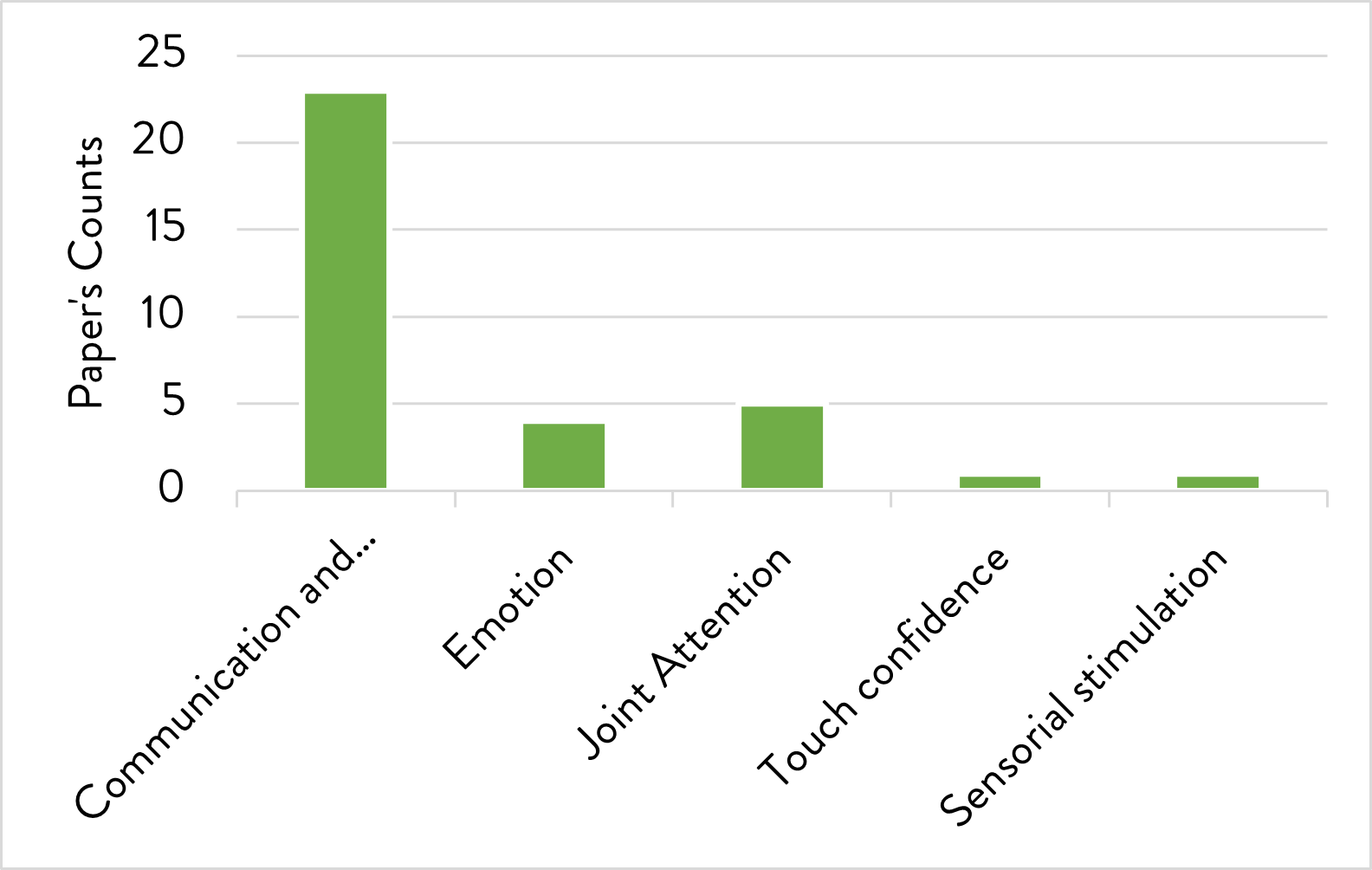}} \subfigure{\includegraphics[width=0.32\textwidth]{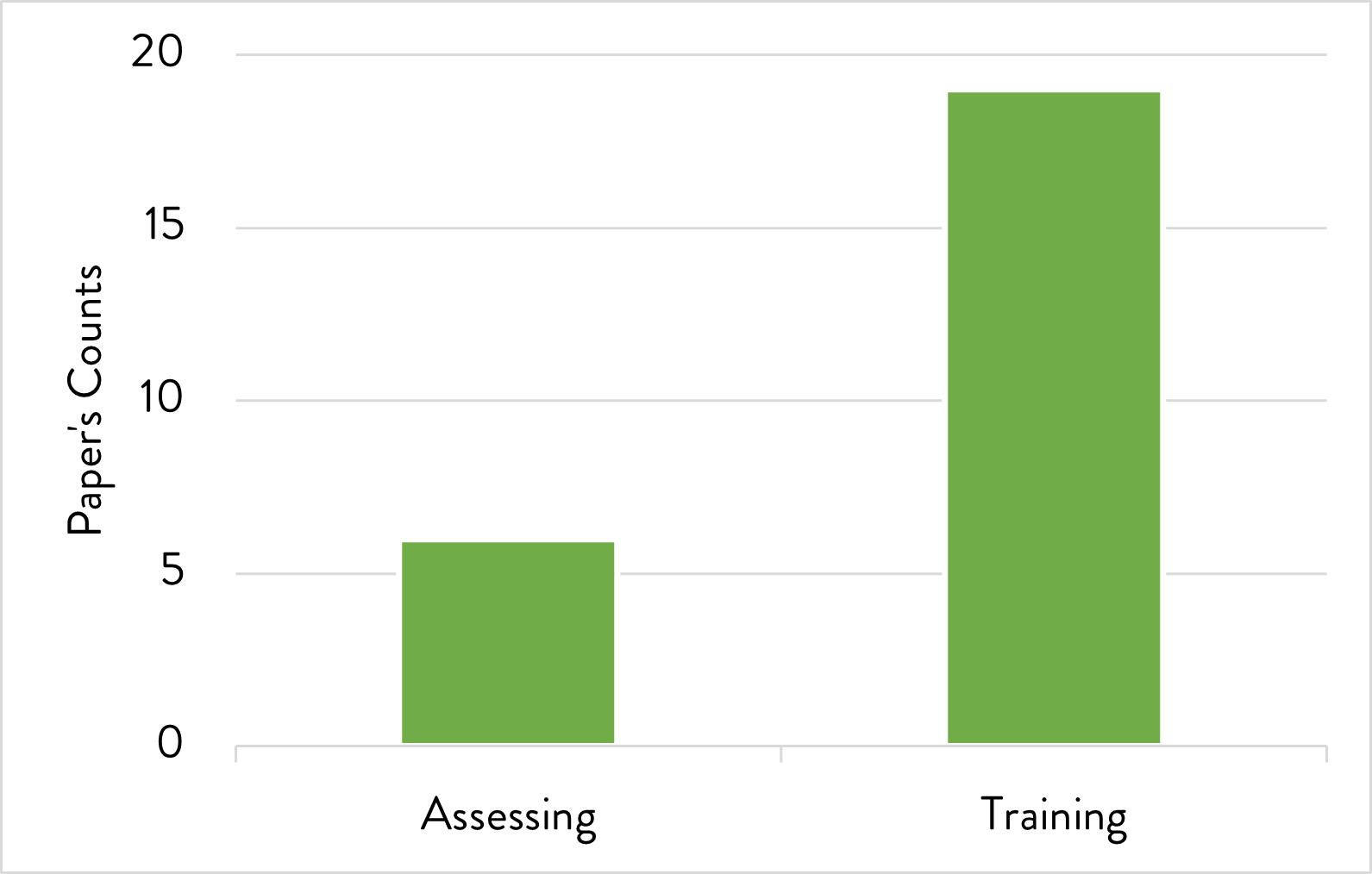}}
\caption{Participant’s skills addressed and agent’s goal (R5, R6): participant's skill addressed (left) and agent's goal (right). }
\label{fig:skillsgoal}
\end{figure}

23 papers (96\%) addressed more than one participant's skill at a time. 
Among the surveyed papers, the authors  addressed emotional skills (4 times in studies, 17\%), joint attention (5 times in studies, 21\%), touch confidence (1 study, 4\%), and sensory stimulation (1 study, 4\%).
23 times (96\%) authors reported they addressed communication and social skills in a broad sense (see Figure \ref{fig:skillsgoal}).

We classified the studies surveyed based on the goal that the authors wanted to achieve using the conversational agent. This classification depends on whether agents aimed at training or assessing participants' skills. 19 studies (76\%) focused on training the target population's skills, while only 6 of them (24\%) investigated their assessment. One study (4\%) addressed both training and assessment (see Figure \ref{fig:skillsgoal}).

To sum up, \textit{results showed that, so far, research has focused mainly on exploring conversational agents for training communication and social skills of people with NDD.}

\subsection{Interaction modalities (R7)}

\begin{figure}
\centering
\subfigure{\includegraphics[width=0.32\textwidth]{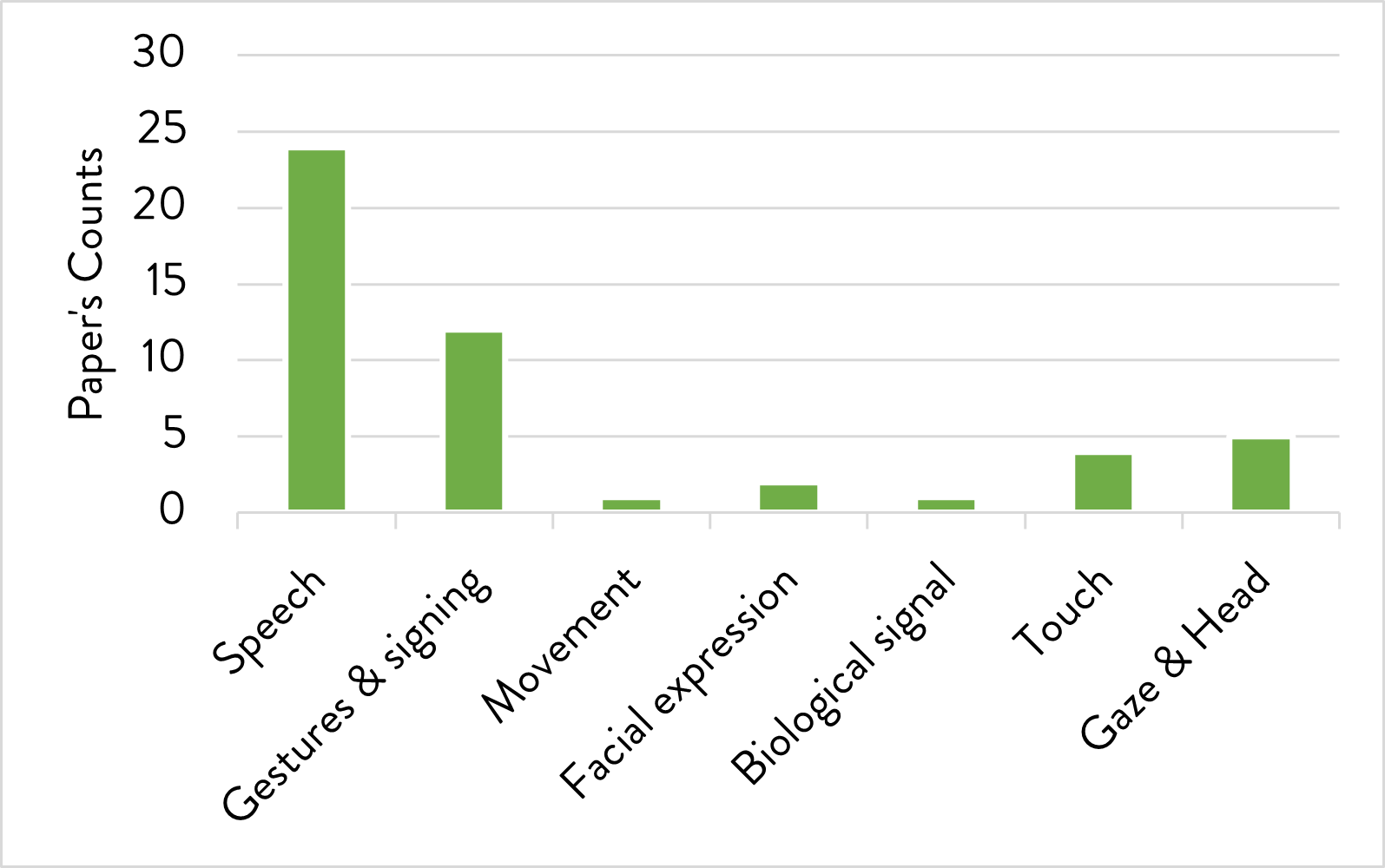}} \subfigure{\includegraphics[width=0.32\textwidth]{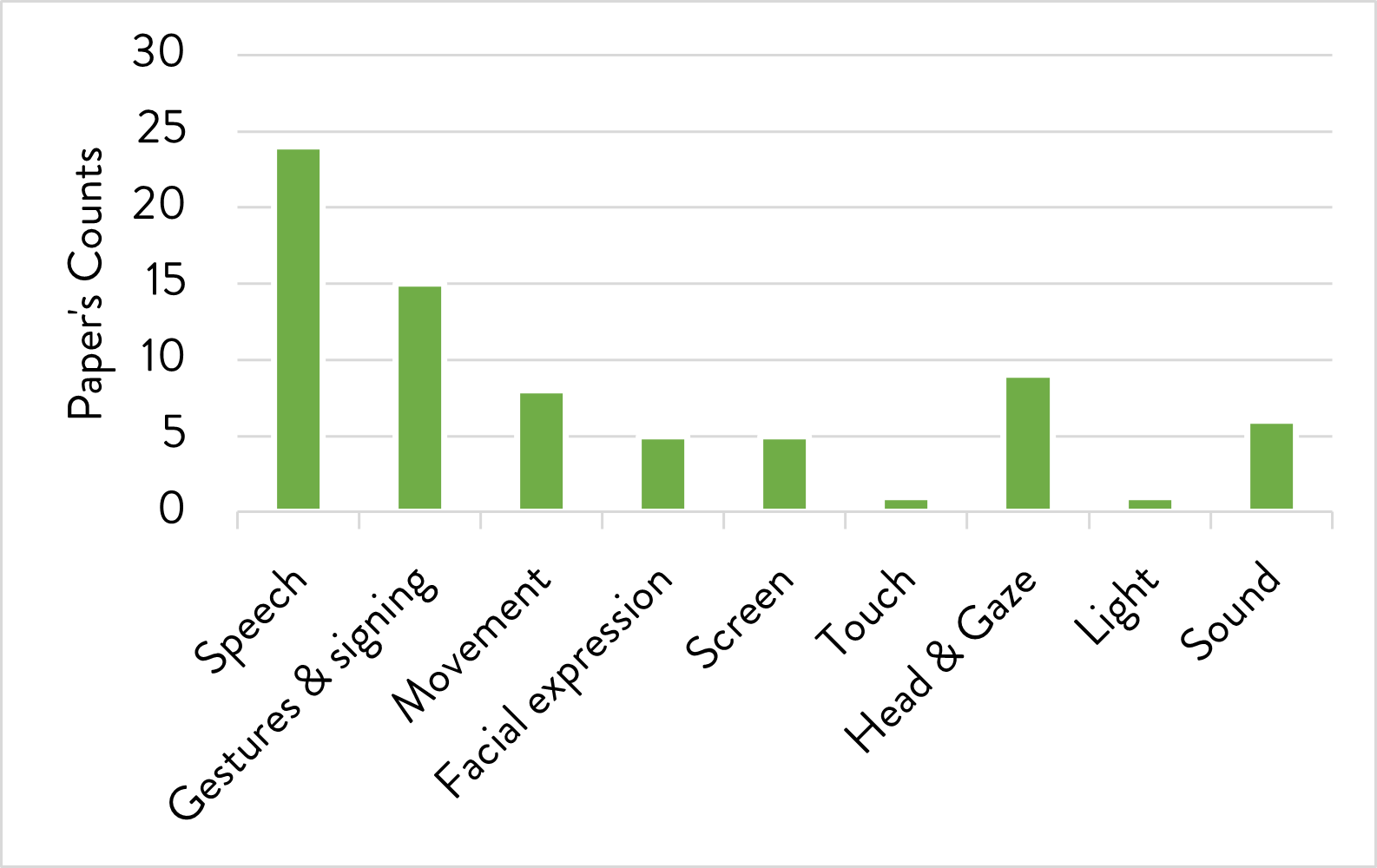}}
\caption{Interaction modalities (R7): input interaction modalities (left) and output interaction modalities (right).}
\label{fig:interactions}
\end{figure}

The conversational agents in the studies we surveyed enabled different interaction modalities (i.e., different types of input and output associated to a specific interaction with a system \cite{karray2017human}), but not all modalities had the same importance: in a number cases, some were fundamental for the experience to take place and others worked just as secondary input or output communication channels between the machine and the user. 
In our analysis, we distinguished between the input interaction modality as the way users communicate with the agent and the output interaction modality as the way the agent communicate with the users.
We noted that both input and output interaction modalities were not always clearly specified in the papers, and it was not always straightforward to extract this information.

All 24 papers (100\%) included speech as input modality. 
12 papers (50\%) described conversational agents able to capture information also from gestures and signs made with the hands by the user.
5 times (21\%) we read about conversational agents able to respond based on user's gaze, 4 times (17\%) based on touch, 2 times (8\%) based on facial expressions, 1 time (4\%) based on parameters coming from biosensors and, finally, 1 time (4\%) based on the identification of the user's movements inside the room (see Figure \ref{fig:interactions}).

Speech was also the primary output modality: 100\% of the papers described conversational agents using speech as communication channel with the user. 
The second most common output modality of interaction for conversational agents was the use of gestures and signs with physical or virtual arms or hands - 15 papers (63\%).
In 9 cases (38\%) the conversational agents used gaze to communicate with the user (sometimes aided by head movement), 8 times (33\%) they managed to communicate by moving through space, 6 times (25\%) they played some sounds, 5 times (21\%) they were able to perform facial expressions, 5 times (21\%) they displayed content on a screen, 1 time (4\%) they used lights as a message to the user, and finally 1 time (4\%) they communicated through the touch (see Figure \ref{fig:interactions}).
To sum up, \textit{our findings showed that the scientific community has employed conversational agents with different input and output interaction modalities for communicating with users with NDD (e.g., agents processing and producing speech and gestures).}

\begin{figure}[!htb]
    \centering
    \begin{minipage}{.32\textwidth}
        \centering
        \includegraphics[width=\textwidth]{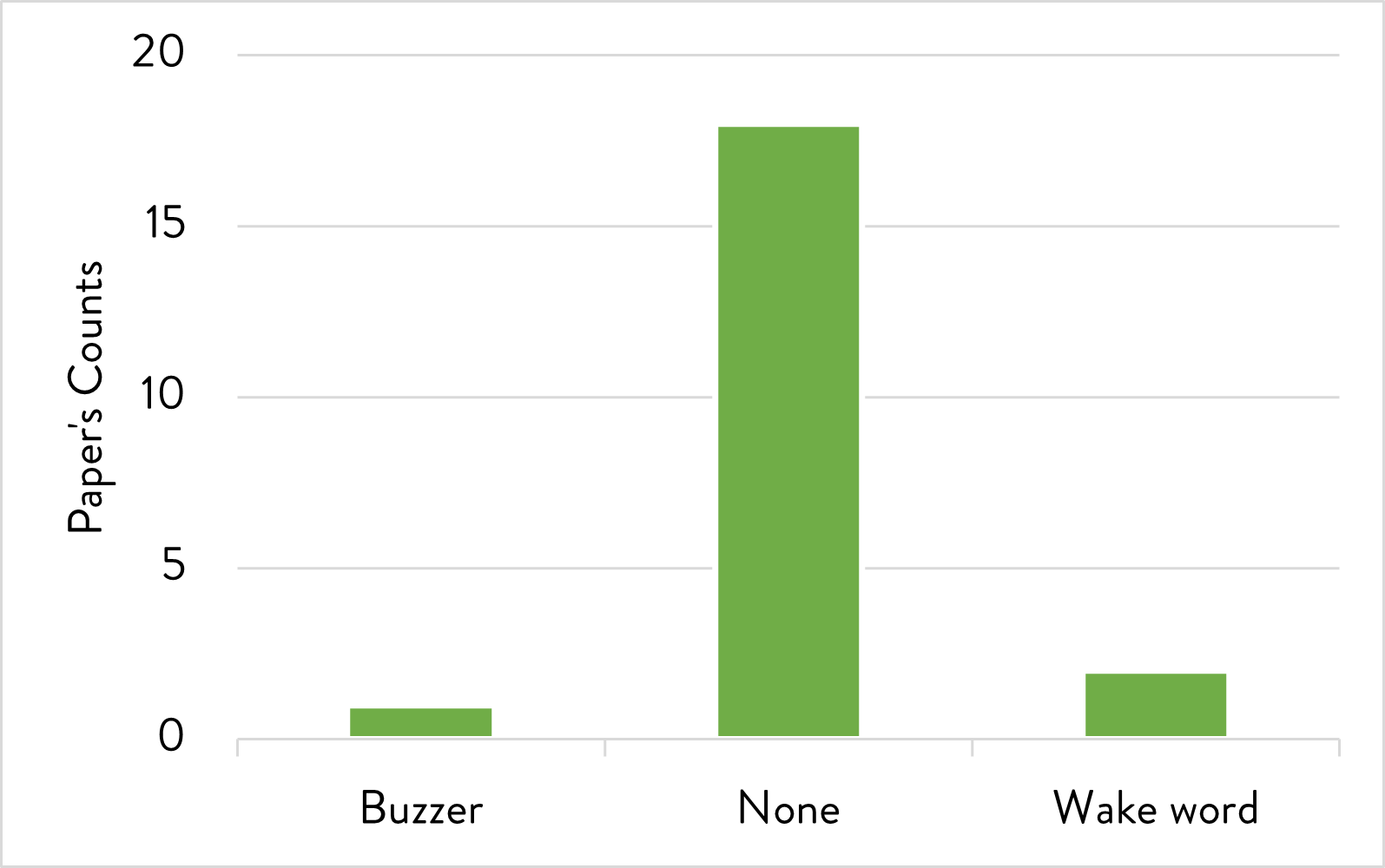}
        \caption{Agent's wake action (R8)}
        \label{fig:wakeaction}
    \end{minipage}%
    \begin{minipage}{.32\textwidth}
        \centering
        \includegraphics[width=\textwidth]{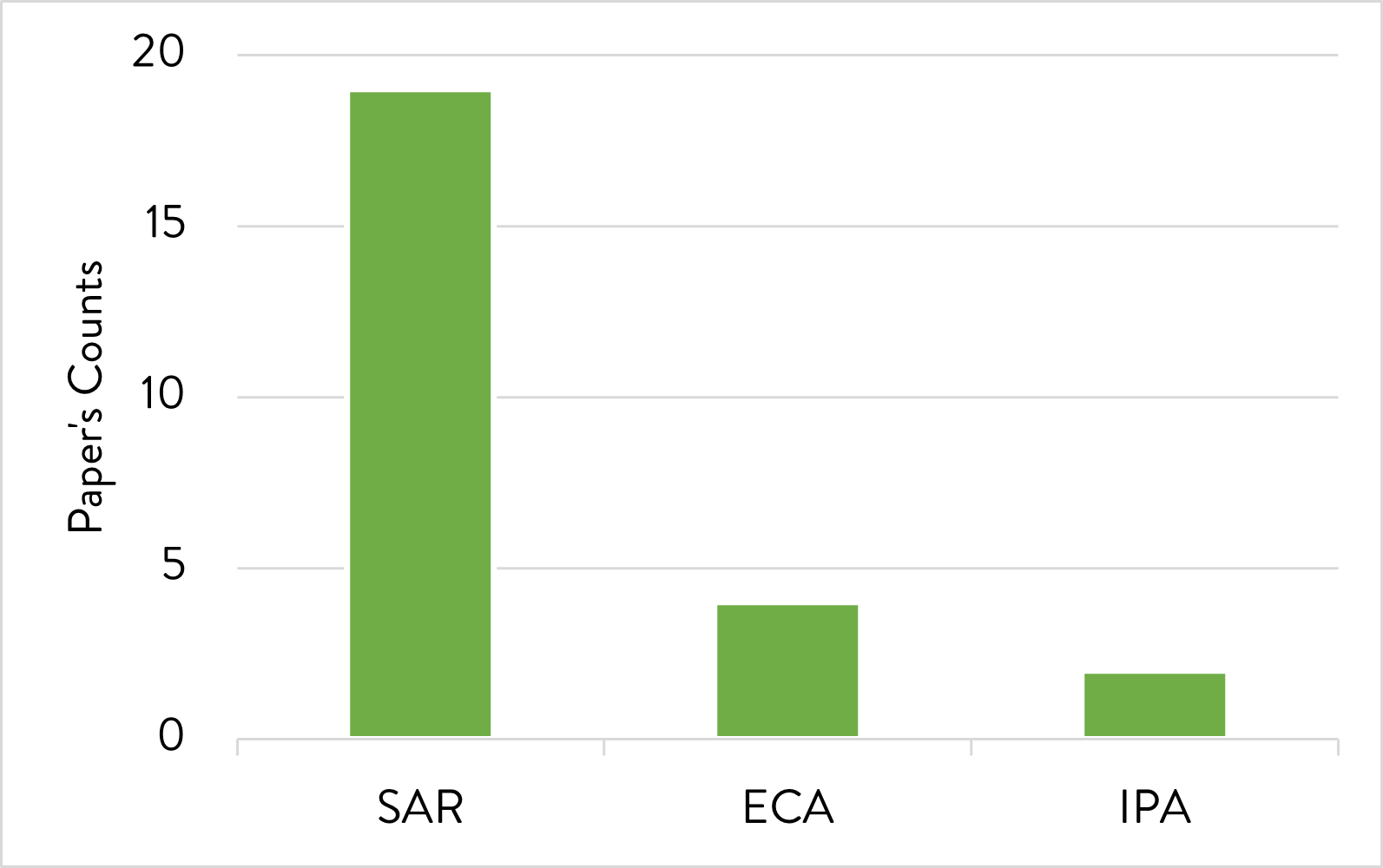}
        \caption{Agent's embodiment (R9). \\ \textit{Legend}: socially assistive robots (SAR), embodied conversational agent (ECA), intelligent personal assistant (IPA).}
        \label{fig:embodim}
    \end{minipage}
    \begin{minipage}{.32\textwidth}
        \centering
        \includegraphics[width=\textwidth]{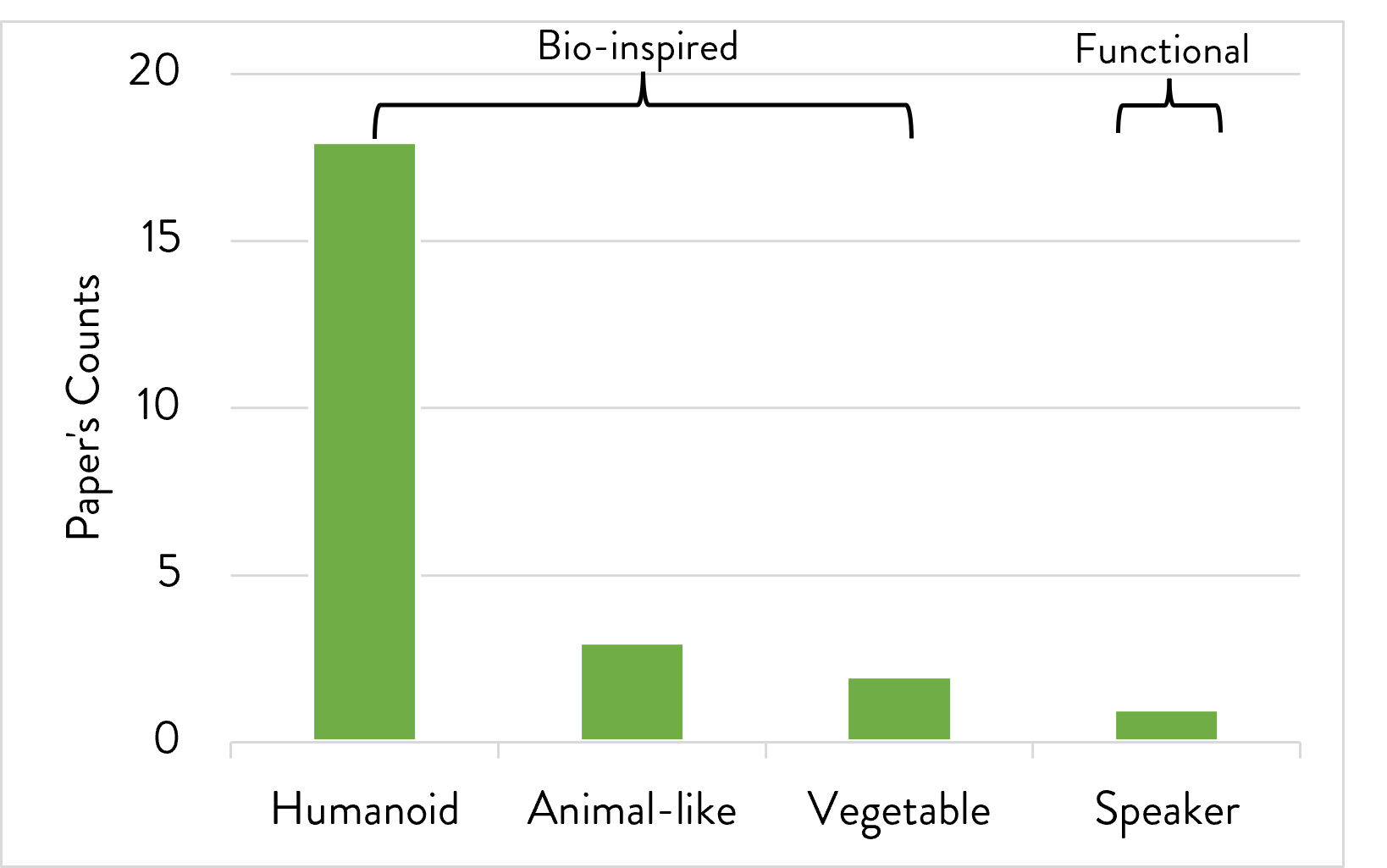}
        \caption{Agent's shape (R10)}
        \label{fig:shap}
    \end{minipage}
     
\end{figure}

\subsection{Agent's wake actions (R8)}
We identified the wake word and the buzzer as wake actions in the surveyed papers. Whenever the authors did not provide any information about it (i.e., none wake action), we assumed that they designed the interaction as a natural human-human communication.  
2 manuscripts (8\%) described conversational agents that used a wake word (e.g., "Alexa" or "Hey, Google") to get activated, while another conversational agent (4\%) got triggered and started listening to the user's words only after playing a beep (event-based \textit{"buzzer"} modality).
The remaining 21 papers in the collection (88\%) described conversational agents functioning without any wake action (none), as reported in Figure \ref{fig:wakeaction}. In those papers, the authors designed the human-agent interaction as a natural speech conversation where the agent is listening to the users respecting turn-taking. 
We noticed that the wake action was specified in all papers describing a conversational agent that implements one. On the contrary, only 3 of the 21 (14\%) papers without a wake action explicitly stated that they did not use one.

To sum up, \textit{results showed that researchers in the field do not usually employ any wake action to trigger the agent at the beginning or during the conversation so as to make human-agent interaction as natural as human-human conversation.}

\subsection{Agent's embodiment (R9)}
19 of the studies on conversational agents for NDD (79\%) employed socially assistive robots (SAR). Embodied conversational agents (ECA) were employed 4 times (16\%) and intelligent personal assistants (IPA) 2 times (8\%). No study included a disembodied conversational agent  (see Figure \ref{fig:embodim}).
Only one study included multiple types of embodiment by considering and comparing a SAR and an ECA. 

To sum up, \textit{results showed that socially assistive robots are the most commonly used agents to interact with people with NDD for therapeutic purposes, followed by embodied conversational agents and intelligent personal assistants.}

\subsection{Agent's shape (R10)}
22 studies (92\%) adopted a bio-inspired shape agent.
Specifically, 18 studies (75\%) employed human-like conversational agents.
The Nao robot by Aldebaran Robotics was employed 10 times, followed by CommU by Vstone Co., Ltd. (2 times), and finally InMoov, Kaspar, and Lucy (1 time). The embodied conversational agent Rachel was used only once as well. 3 studies (12\%) employed animal-shaped conversational agents.
\citet{soleiman2014roboparrot} used the Squawkers McCaw robot shaped as a parrot from Hasbro Toy Company, \citet{spitale2020whom} used an avatar and the Bluetooth Buddy Speaker Lamb robot by iLive shaped as a sheep, and \citet{pop2013social} used the Probo robot that is shaped as an imaginary animal with elephant-like features.
Only one study (4\%) employed a flower-shaped conversational agent, namely the socially assistive robot Daisy.
The only two studies on IPAs (8\%) opted for a functional shape using the off-the-shelf Alexa integrated into Amazon Echo Dot and Google Assistant integrated into Google Home. Both devices are in the form of speakers.
Figure \ref{fig:shap} depicts those shapes.

To sum up, \textit{our findings showed that the scientific community has explored functional shapes for conversational agents for NDD but normally leans towards the choice of bio-inspired shapes, specifically with a human-like appearance.}

\begin{figure}[!htb]
    \centering
     \begin{minipage}{.32\textwidth}
        \centering
        \includegraphics[width=\textwidth]{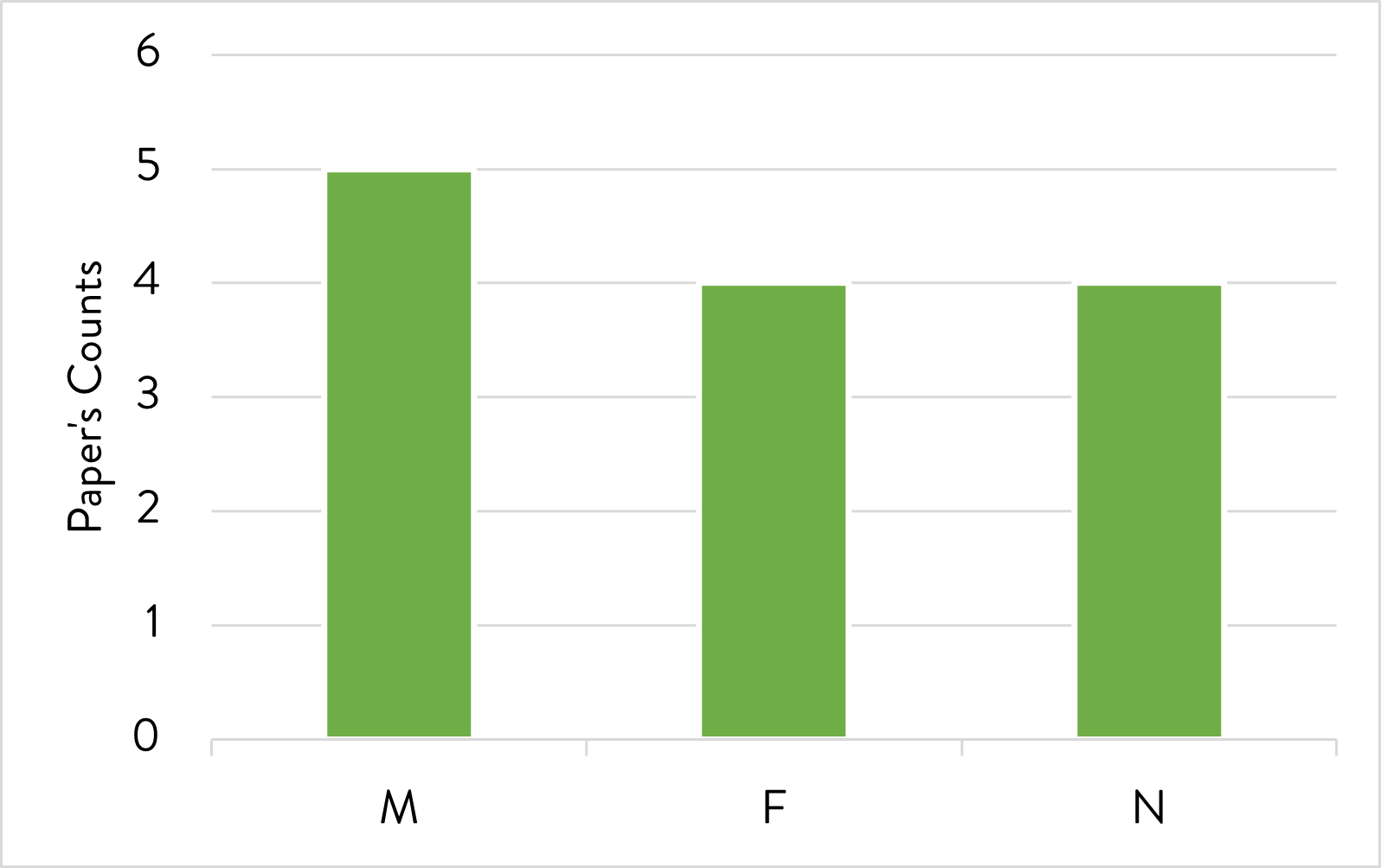}
        \caption{Agent's gender (R11). \\\textit{Legend}: male (M), female (F), neutral (N).}
        \label{fig:gend}
    \end{minipage}%
    \begin{minipage}{.32\textwidth}
        \centering
        \includegraphics[width=\textwidth]{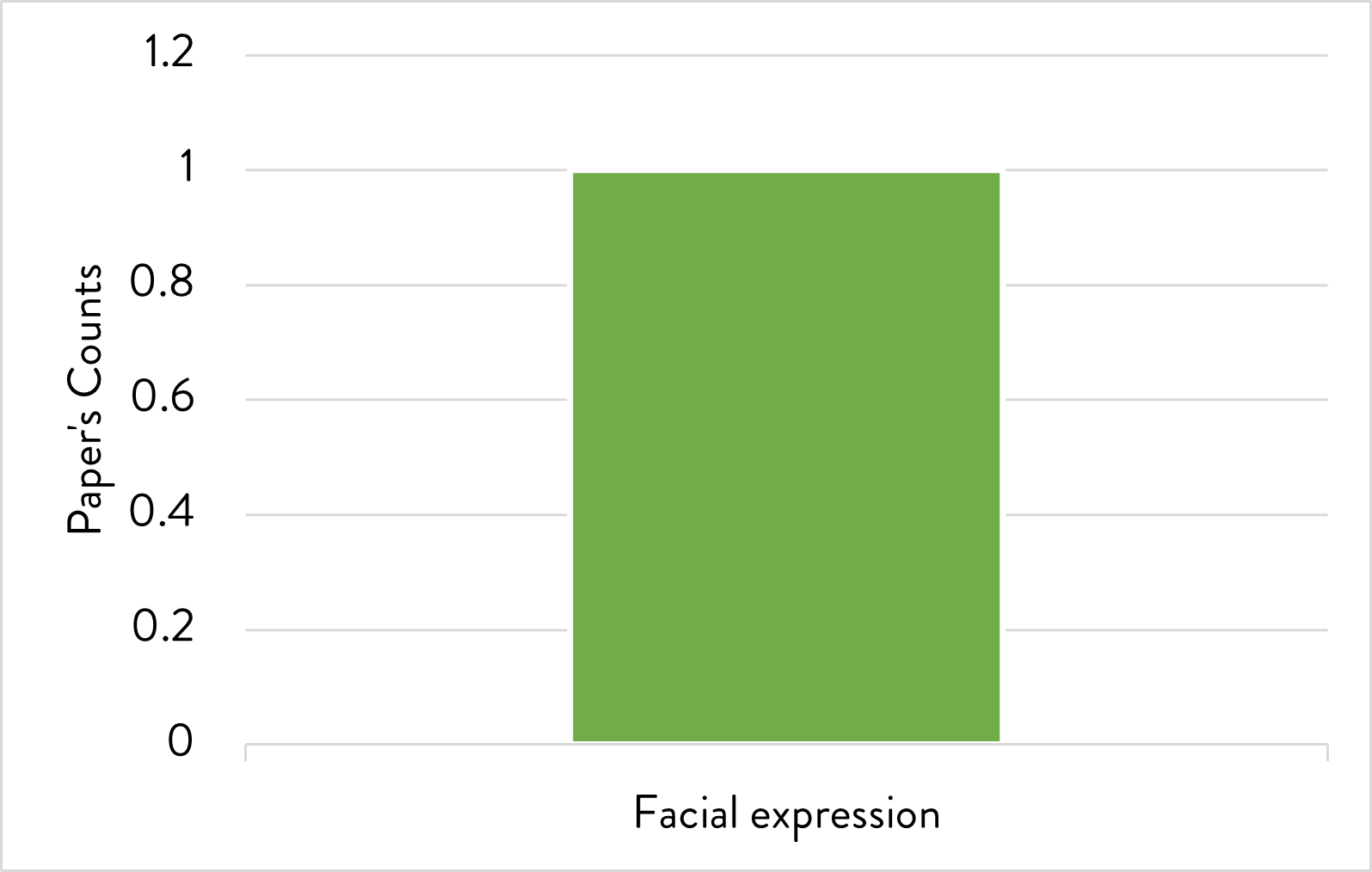}
        \caption{Agent's emotional recognition capabilities (R12).}
        \label{fig:emot}
    \end{minipage}
    \begin{minipage}{.32\textwidth}
        \centering
        \includegraphics[width=\textwidth]{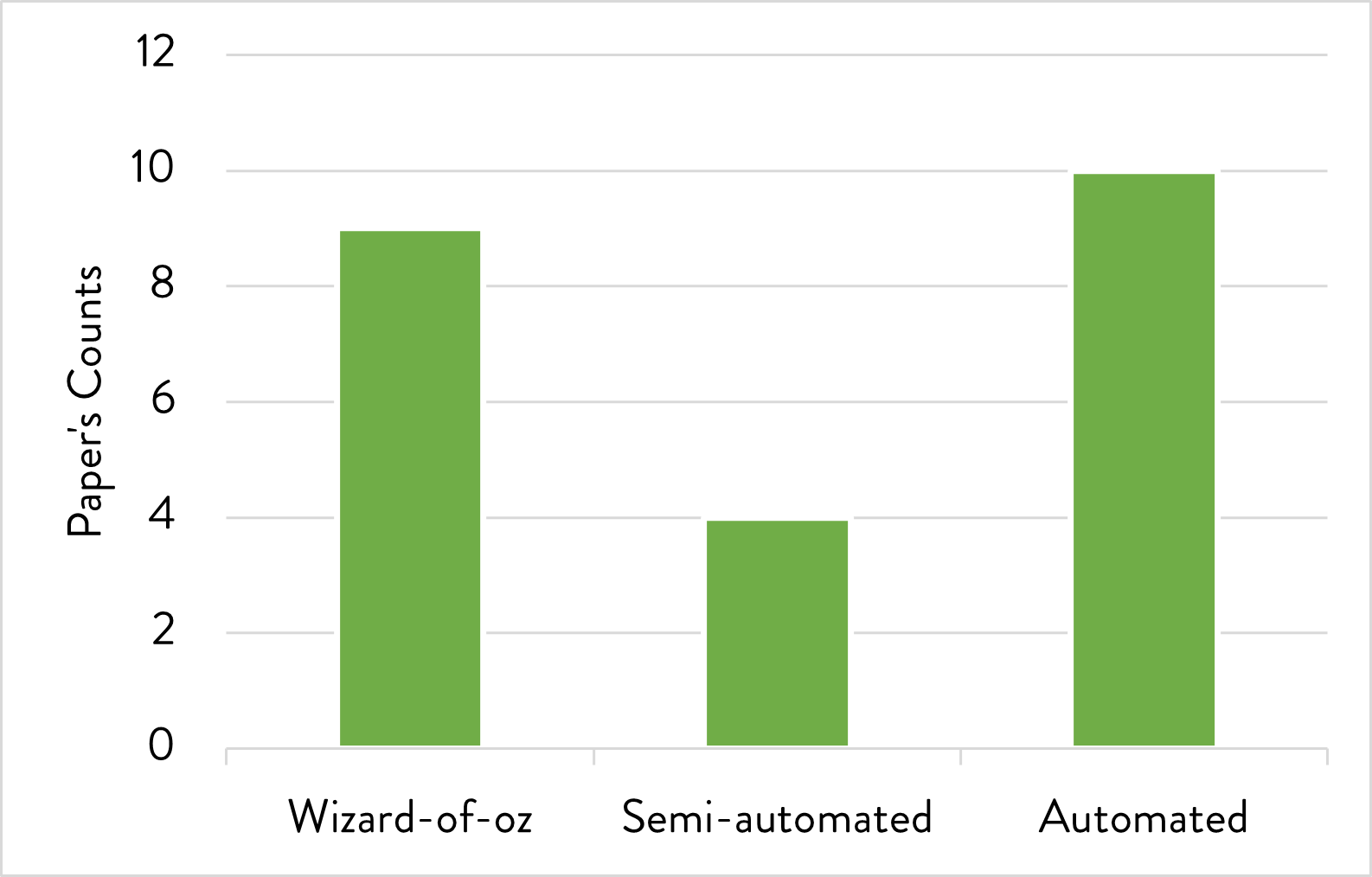}
        \caption{Nature of the prototype (R13)}
        \label{fig:prot}
    \end{minipage}
\end{figure}

\subsection{Agent's gender (R11)}
About the gender of the conversational agents (see Figure \ref{fig:gend}), 4 studies (17\%) employed a male agent.
3 studies (12\%) used female agents and other 4 studies (17\%) employed neutral agents. 1 study (4\%) used both female and male agents.
We report that 12 studies (50\%) did not specify the gender of the conversational agent and we could not infer it in any way not even from the images in the manuscript.
The gender of a conversational agent can be made explicit in several ways, from appearance to voice. Neutral agents are achieved through an appearance that is explicitly neither masculine nor feminine and with a synthesized voice that can be robotic or non-gender specific. 
For example, \citet{axelsson2019participatory} chose not to wear the robot InMoov with any clothes both to make it more simple looking and gender-neutral. Also, they modified the text-to-speech engine: a female human-like voice was used, and its pitch lowered to make it neutral. The neuropsychologist advocated for a robotic voice, since the children might confuse the robot with a human if it were too lifelike. Consequently, the voice of the robot was slowed by 10\%.

To sum up, \textit{results showed that, despite NDD experts seek for neutral-gendered agents, the current research has adopted male-, female- and neutral-gendered agents without major differences.}

\subsection{Agent's emotional recognition capabilities (R12)}
We researched whether conversational agents for people with NDD were developed with capabilities to recognize users' sentiment or emotions. We found out that only one study (4\%) described implementing such a functionality (see Figure \ref{fig:emot}).
The socially assistive robot used by \citet{khosla2015service} exploited a module for sentiment recognition from facial expressions to customize the interaction in real-time based on the detected mood of the user. Depending on the result of the sentiment analysis, the conversational agent adapted the responses to the user in terms of verbal and non-verbal cues.
Other authors (e.g., \cite{tanaka2017embodied, bekele2016multimodal}) reported about conversational agents able to extract audiovisual features during the interaction with the user to analyze their social cues. 
In the case of \citet{tanaka2017embodied}, for example, features extracted referred to user’s speech, pitch, language, smiling ratio, and yaw and enabled the agent to adapt the interaction and provide different feedback depending on them. In all these cases, one may still consider features as markers of an emotional state of the user, but they were actually used directly without running any sentiment or emotional analysis.

To sum up, \textit{results showed that endowing the conversational agent for NDD with sentiment and emotional capabilities is still an unexplored feature.}

\subsection{Nature of the prototype (R13)}
During their empirical experimentation, 9 studies (39\%) exploited a wizard-of-Oz prototype, 4 of them (17\%) adopted the semi-autonomous agent, and 10 of them (43\%) an autonomous prototype (see Figure \ref{fig:prot}).
For one study (4\%) we were not able to understand the nature of the prototype because authors did not provide enough information.

To sum up, \textit{our findings showed that, despite the interest in autonomous agents increased during the last years, many researchers still exploit the wizard-of-Oz approach to investigate human-computer interaction aspects (e.g., user's fatigue and perception).}

\begin{figure}[!htb]
    \centering
    \begin{minipage}{.32\textwidth}
        \centering
        \includegraphics[width=\textwidth]{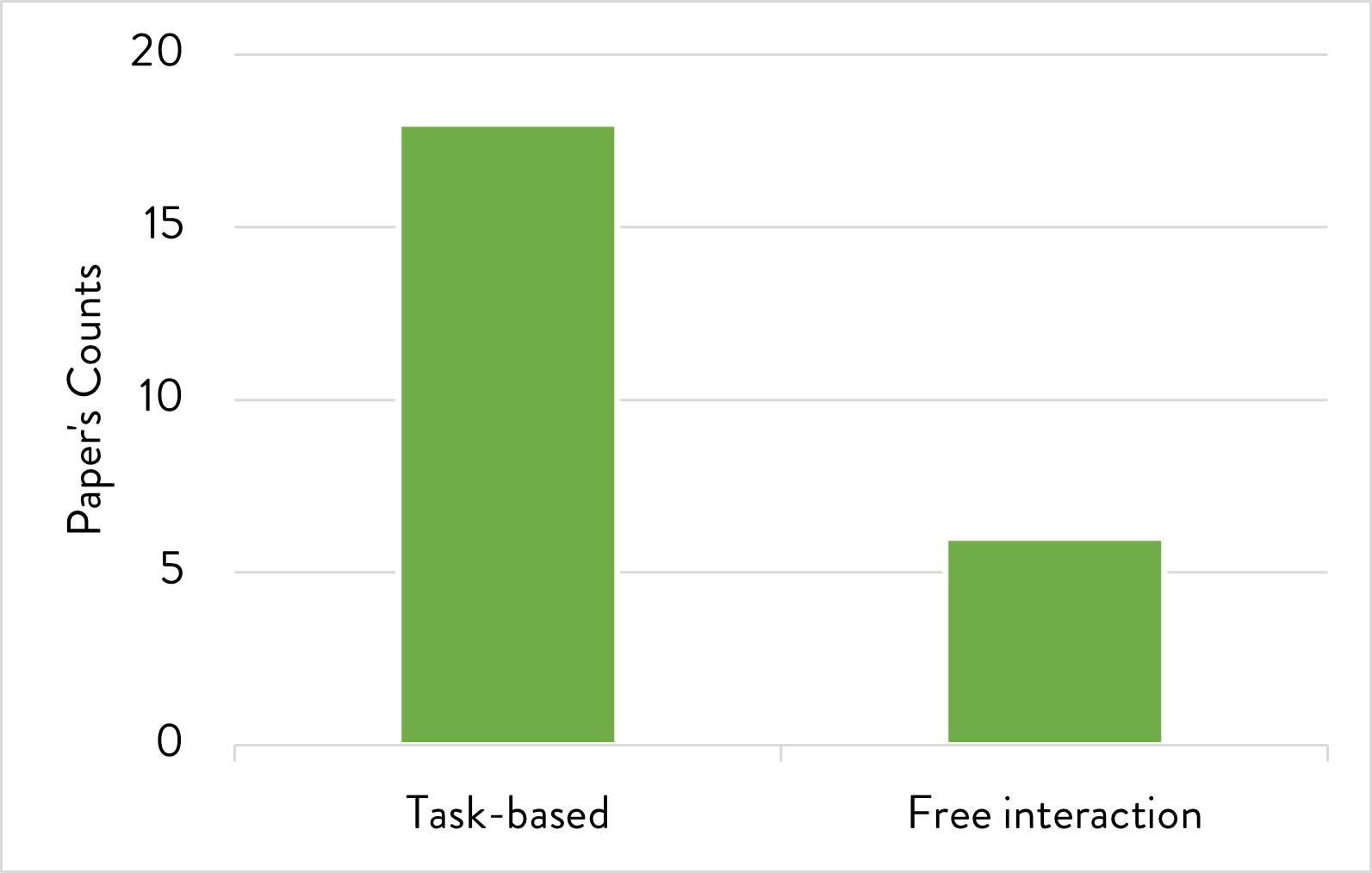}
        \caption{Intervention type: task-based vs. free interaction (R14)}
        \label{fig:interv}
    \end{minipage}%
    
\end{figure}

\subsection{Intervention type: task-based vs. free interaction (R14)}
We categorized the intervention types as in \cite{kabacinska2021socially}, where authors identified task-based (i.e., structured and guided interaction) and free interactions (i.e., unstructured interaction). 
The majority of the surveyed studies (18 - 75\%) adopted task-based activities during the sessions with the agent to assess or train a specific skill of the user (e.g., proposing a memory game to train users' memory skills).
Only 6 of them (25\% of the studies) asked participants to freely interact with the conversational agent without providing them any specific instructions or any task to accomplish to explore the interaction in a more natural and ecological context (see Figure \ref{fig:interv}).

To sum up, \textit{our findings showed that the researchers have designed conversational agents for NDD mainly for performing task-based activities rather than free interaction.}

\begin{figure}
\centering
\subfigure{\includegraphics[width=0.32\textwidth]{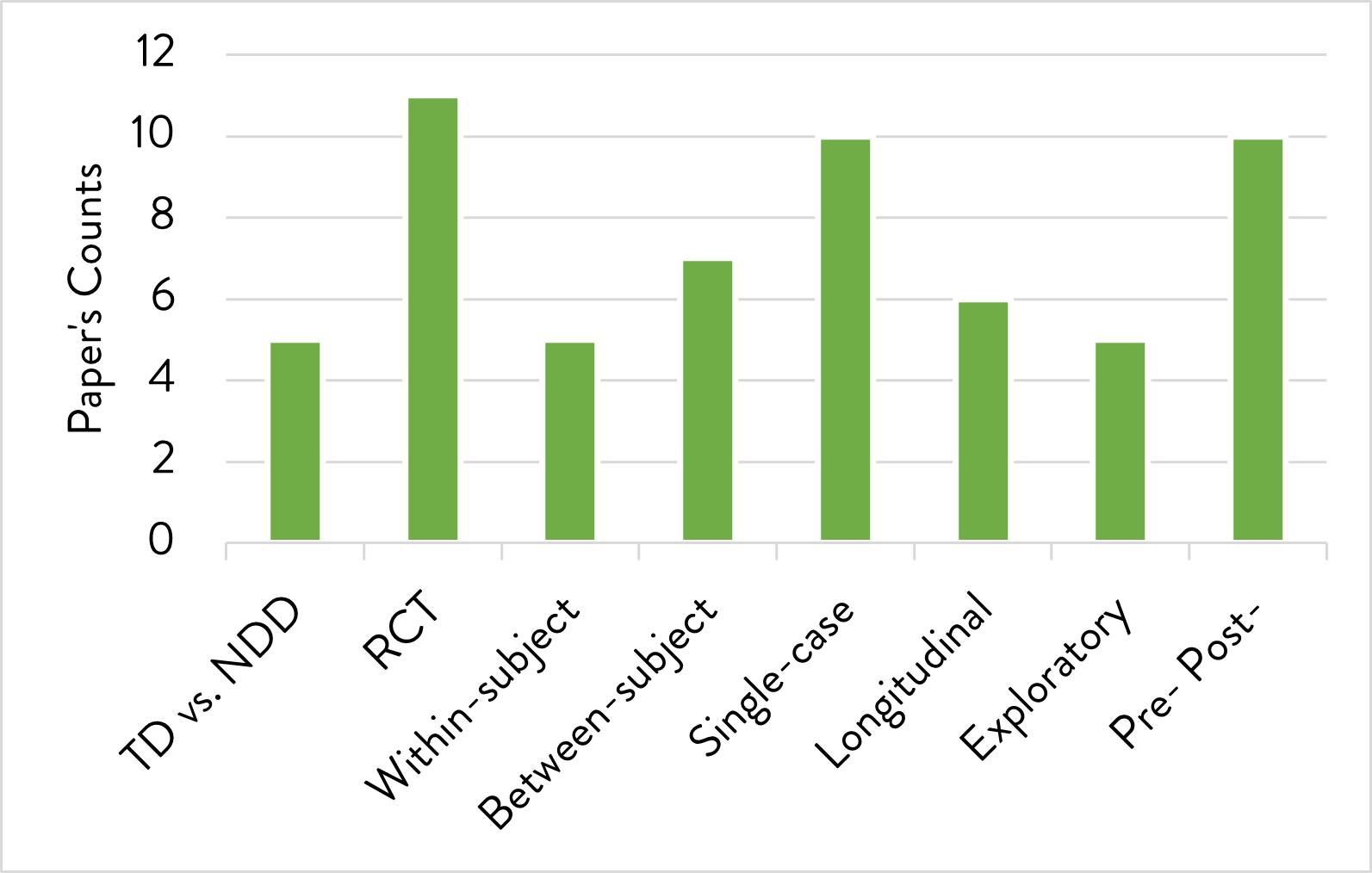}} \subfigure{\includegraphics[width=0.32\textwidth]{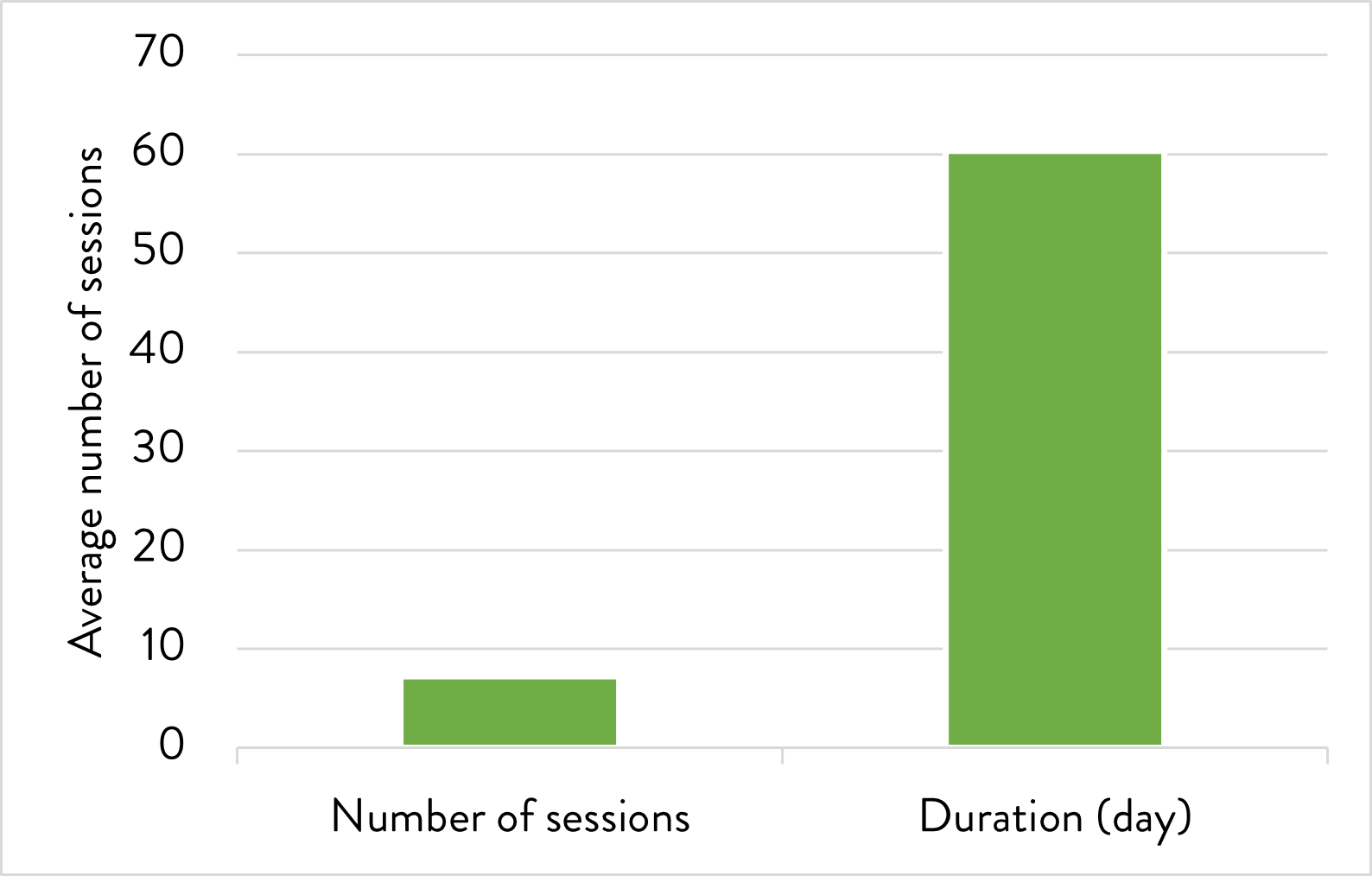}}
\caption{Study design and duration (R15, R16): study design (left) and study duration (right).\\ \textit{Legend}: typically-developed (TD), random control trail (RCT).}
\label{fig:study}
\end{figure}

\subsection{Study design and duration (R15, R16)}
To frame the methodology of their studies, the authors of the surveyed papers chose among different well-known study designs in literature depending on the purpose of their research and the sample size they recruited (see Figure \ref{fig:study}). 
11 studies (46\%) adopted the Randomized Controlled Trail (RCT) design where researchers randomly assigned participants to groups, for example to a control or experimental condition. Among those, 5 of them assessed the skills addressed pre- and post- training to evaluate participants' improvement after the intervention with the agent. In total, 10 studies surveyed (42\%) used the pre- and post- approach to assess the participants improvements after training. 
Other 5 studies (21\%) involved both the TD and NDD populations to compare people with NDD behavior to the conventional population (TD) and to understand the benefits of using a conversational agent with one sample with respect to the other. 7 studies (29\%) used a between-subject study (i.e., each participant has been assigned to one of two or more conditions) design and other 5 (21\%) a within-subject study design (i.e., each participants experience all the conditions designed for the study). 9 out those 13 studies (69\%), with either a between- or within-subject study, adopted also the RCT technique to avoid bias randomly assigning their participants to the study conditions. 
6 studies (10\%) had a longitudinal design, that means that their studies lasted for several weeks, in some cases even months. 
Other 5 studies (25\%) were exploratory without any specific design.
10 studies (42\%) had a single-subject design because they did not have a sample large enough to draw conclusion about the general population. 

The duration of the studies surveyed was on average 60 days (SD = 79.66), as reported in Figure \ref{fig:study}. The standard deviation is so high because some of the studies were single session and others were very long longitudinal studies. Specifically, 7 studies (30\%) exploited a single-session study, 11 of them (48\%) experimented among 2 and 9 sessions, and only 5 (22\%) lasted for more than 10 sessions (see Figure \ref{fig:study}).

To sum up, \textit{results showed that, so far, research in the field has adopted mostly a RCT design approach and a single-subject study design, and empirical investigation generally lasts no more than 9 sessions in two months time.}

\begin{figure}[!htb]
    \centering
    \begin{minipage}{.32\textwidth}
        \centering
        \includegraphics[width=\textwidth]{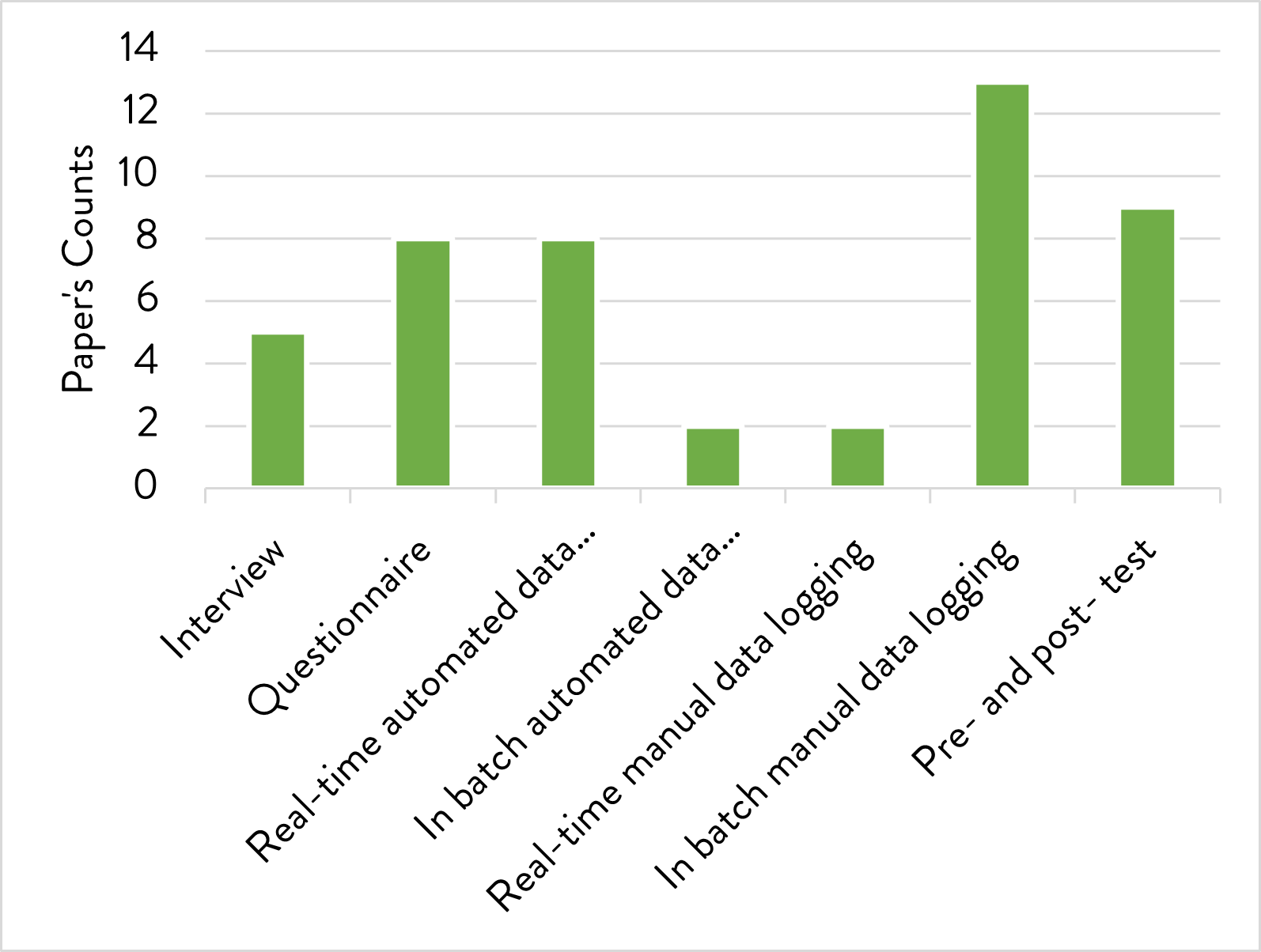}
        \caption{Methods of data collection (R17)}
        \label{fig:datacoll}
    \end{minipage}%
    
\end{figure}

\subsection{Methods of data collection (R17)}
We identified seven research methods that have been adopted to collect data during the empirical studies surveyed (see Figure \ref{fig:datacoll}). 
In 13 studies (34\%) researchers manually annotated the collected data in batch, in 8 of them (21\%) they asked participants to fill questionnaires (most of them adopting a Likert scale to indicate their level of agreement with the questionnaire's items), 8 studies (21\%) collected real-time data with an automatic logging, in 5 studies (13\%) the authors interviewed the participants to gather qualitative data, 2 of them (5\%) took note manually in real-time during the sessions with the users, while other 2 studies (5\%) exploited a software to perform automated data logging from video and audio recordings in batch. 

To sum up, \textit{results showed that researchers mostly exploit either subjective (e.g., interview, questionnaires) and objective (e.g., analyzing video and audio recordings) data collection methods and usually annotate it manually and in-batch}.

\subsection{Study's findings (R18, R19, R20)}
\subsubsection{Methodology}

Within the studies surveyed, only the authors in \cite{Huijnen2021-nx} highlighted a methodological result. They ran a 4-sessions ABAB empirical study with 9 children with autism (2 session with the KASPAR robot, and 2 with the teacher) to train their communication and social skills. They collected quantitative and qualitative data via respectively video recordings of the interaction and via interview with teachers. To analyze the video recordings of the interaction, they used a dialogue coding scheme approach where they labeled children's micro-behaviors classifying them into verbal (e.g., positive verbal utterance on initiative child) and non-verbal (e.g., increase of physical proximity). 
They combined the results from the analysis of children's micro-behaviors (quantitative data) and the results of the interview with the teacher (qualitative data).
Their findings showed that KASPAR promoted making contact with children with autism more than the teacher did, highlighting the potential of socially assistive robots to improve communication and social skills. Those results emerged from both quantitative and qualitative analysis. Specifically, their mixed-method approach (i.e., quantitative and qualitative) showed in-line findings that confirm each other. 

To sum up, \textit{our findings showed that mixed-method approaches, which consider both quantitative and qualitative data, are a powerful method to obtain stable and robust results from empirical studies.}

\subsubsection{Technology}
Only one study (4\%) included findings related to technology. 
\citet{Cao2019-yg} addressed the problem of measuring participants' joint attention from analyzing the video recordings of their sessions. 
Tracking the attention level of people with NDD is important since it can be used for estimating users' cognitive skills and for assessing the quality of their interaction with technology.
Previous studies relied on fixation time, which is the sum of the time spent by the participant on looking at the area of interest. In this case, the authors proposed using the longest common subsequence (LCS) algorithm, which is commonly and widely used in computational biology and human genome sequencing. Using the LCS algorithm, the similarity between the eye movement sequences of participants and a reference sequence can be calculated by traversing the two sequences. 

To sum up, \textit{results showed that the longest common subsequence (LCS) algorithm is a good tool for measuring joint attention during human-agent interaction and can be potentially applied to much more complex situations than the standard fixation analysis.}

\subsubsection{Design}
In this sub-subsection, we grouped the papers according to the skills that the authors wanted to assess or train (see subsection \ref{goalandskills}) and we reported what features were chosen for designing the conversational agents addressing each user's skill.

\paragraph{Emotional skills}
Out of the four studies addressing emotional skills, 2 (50\%) employed socially assistive robots, and the others (50\%) used embodied conversational agents. 3 studies (75\%) employed a human-shaped agent, and 1 study (25\%) used a flower-shaped one. 2 papers (50\%) did not specify the gender of their conversational agents, while the other 2 papers employed three agents in total, two females and one male. 
Beyond speech, 2 agents (50\%) accepted input in the form of gestures, gaze, touch, and data from biosensors. 4 of them (100\%) produced output in the form of speech, 3 (75\%) used gaze, 2 (50\%) facial expressions, 2 (50\%) gestures, 1 (25\%) produced visualizations on the screen, and 1  (25\%) could move through the space. None of the agents implemented a wake action, and none could recognize the user's emotions. 

\citet{mower2011rachel} designed a woman-like embodied conversational agent called Rachel that acted as an emotional companion for autistic children to create semantically emotional narratives. 
After validating the system with two children with ASD, early results suggested that Rachel offered a viable means for collecting targeted socio-emotional interaction data to assess the emotional skills of the users. 
Similarly, \citet{marino2020outcomes} employed a Nao humanoid-shaped socially assistive robot 
to support a cognitive behavioral therapy intervention for 14 children with autism in training their emotional skills. 
Their results showed significant improvements in contextualized emotion recognition, comprehension, and emotional perspective-taking were attained through the use of Nao.
Those findings suggested that a conversational agent could be successfully used for assessing and training the emotional skills of children with autism.
\citet{bekele2016multimodal} presented two virtual reality-based human embodied conversational agents 
who aimed to stimulate the emotional skills of children with autism via an emotion recognition tasks. 
A between-subject usability study investigated the performance of 12 children with ASD. Half of them (experimental group) interacted with a gaze-sensitive version of the system 
and the other half (control group) interacted with the system without the gaze adaptation. 
The authors suggested that the gaze-sensitive system enabled the former group to improve better their emotional skills over time than the control group. Those finding suggested the value of using social cues to adapt the interaction to improve further the children's emotional skills. 
\citet{fachantidis2020tauhe} compared the interaction - eye
contact, proximity and verbal interaction - between four children with autism spectrum disorder and an agent who could be a human being or a flower-shaped robot, called Daisy. 
During the sessions, the agent proposed four activities to participants and sought to train users' understanding and comprehension of emotions.
Results indicated that children who face difficulties in human-human interactions smoothly communicated with the robot since it manifests emotional expressions in an easily recognizable way.
Their findings highlighted the potential of exploiting the emotional training activities with robots to promote the communication of children with autism.

To sum up, \textit{results showed that both socially assistive robots and embodied conversational agents could provide an effective platform for training the emotional skills of people with NDD and encouraging affective and social behavior. Also, real-time elicitation and analysis of the user's social cues (e.g., the gaze) provided helpful information about the ongoing interaction. That kind of information could be used to adapt the agent's behavior and positively impact the effectiveness of the therapeutic intervention in general.}

\paragraph{Joint attention}
5 studies addressed joint attention skills, and each of them employed a human-shaped socially assistive robot. Just one paper (20\%) used a neutral agent, while the other 4 papers (80\%) did not specify the gender of the conversational agent. Apart from speech, 3 agents (60\%) accepted input in the form of gestures, and 1 agent (20\%) used gaze.
5 agents (100\%) produced output in the form of speech, 5 (100\%) used gestures, 2 (40\%) moved through the space, 1 (20\%) used gaze, 1 (20\%) used light, and 1 (20\%) could play some sounds. 4 agents (80\%) did not implement any wake action, while the last (20\%) was triggered only after playing a beep. No agent could recognize the user's emotions.
Next, we describe the design results of the surveyed paper which conversational agent aimed to promote the joint attention of the target population.

\citet{Cao2019-yg} used a Nao robot. The authors investigated whether the interaction with the robot could promote joint attention in children with ASD. Their study involved 15 typical-development children and 15 children with ASD, and they measured the joint attention to the robot and human conditions. Their early-stage findings showed that children with ASD showed more interest in the robot's face, promoting more their joint attention then in the human condition.
Similarly, \citet{axelsson2019participatory} employed a InMoov robot to 
interact with ten children with ASD for imitating assistive sign language (i.e.,  the most common form of assistive and augmentative communication used by people with ASD \cite{von2000introduction}). 
Results showed that children with ASD successfully imitated the robot and kept their attention throughout the interaction. 
Analogously, \citet{So2020-oq} used two Nao robots 
to investigated whether a robot-based play-drama intervention could enhance joint attention and play behaviors of children with autism. Twelve children with autism interacted with the robot (intervention group), and 11 children with autism interacted with a human (control group) for nine weeks. Findings showed that children of the intervention group significantly improved their joint attention and played behavior compared with the control group. 
\citet{amanatiadis2017interactive} employed a Nao robot 
during a four-session study where children with mild ASD were asked to imitate the words and movements of the robot. Findings showed that children paid more attention to the robot and could respond to a statement/gesture issued by the agent more readily than to a statement/gesture issued by a human. 
Similarly, \citet{Van_Otterdijk2020-ww} used a Nao robot 
to investigate whether the interaction with the robot changed in quality over time. They ran a study with six children with autism over 20 sessions focused on training their non-verbal behaviors. Results showed that in the post-intervention session, children had increased sustained attention and engagement with their parents. 

To sump up, \textit{results showed that humanoid socially assistive robots have the potential to improve joint attention of people with NDD.}

\paragraph{Touch confidence}
Only one paper addressed the improvement of touch confidence, meant as the ability to make contact with a person.
\citet{Huijnen2021-nx} tried to assess the touch confidence of children with ASD through a conversational agent. They employed a male humanoid robot, namely KASPAR. 
The authors involved nine children in a four-session mixed-method study with ABAB design (two sessions with the agent and two with the teacher). Results showed that in the KASPAR condition children searched the physical contact more than in the teacher condition.

To sum up, \textit{results showed that a humanoid robot can boost the physical contact of children with autism.}

\paragraph{Sensorial skills}
Again, only one paper addresses the sensorial skills' problem.
\citet{catania2021toward} sought to address the sensorial skills of children with ASD with conversational technology. They employed Google Home, which is a male intelligent personal assistant shaped like a speaker. It implements the wake word to activate (i.e., "Hey Google"), understands the user's speech, and produces speech and sounds.
The authors conducted an exploratory study that investigated the use of the IPA in a therapeutic setting with nine children with NDD. They provided three therapists with Google Home for twenty-one days, and therapists were welcome to use the IPA as they liked at their ordinary one-to-one therapy sessions. Findings showed that Google Home could stimulate and reward children with high severity in speech-based communication and socialization. However, since children's cognitive, sensory, and language impairments make it difficult for them to autonomously use Google Assistant, the interaction with the agent should be facilitated by a therapist.

To sum up, \textit{results showed that a conversational agent such as Google Home has the potential for being used as a sensorial stimulus mainly for children with high-severity NDD because it enables speech-based interaction for playing some sounds as part of the activities and as a reward, reinforcement, and relaxation tool}.

\paragraph{Communication and social skills in general}
23 studies of the survey stated they addressed communication and social skills in general.
18 (78\%) employed socially assistive robots, 4 (17\%) used embodied conversational agents, and 2 (9\%) explored intelligent personal assistants. 17 (74\%) studies used humanoid-shaped agents, 3 (13\%) employed animal-shaped agents, 2 (9\%) used speaker-shaped agents, and 1 (4\%) worked with a flower-shaped agent.
5 papers (22\%) employed a male agent, 4 (17\%) used a female agent, and 4 (17\%) employed a neutral agent. 11 papers (48\%) did not specify the gender of the agent.
Apart from speech, 12 agents (52\%) accepted input as gestures, 5 (22\%) as the gaze, 4 (17\%) as touch, 2 (9\%) as facial expressions, 1 (4\%) as data from biosensors, and 1 (4\%) in the form of movements in space.
23 studies (100\%) involved agents producing output in speech, 14 (61\%) in the form of gestures, 8 (35\%) as the gaze, 6 (26\%) as sound, 7 (30\%) moving in space, 5 (22\%) as facial expressions, 5 (22\%) as visualizations on a screen, 1 (4\%) using touch and 1 (4\%) using lights.
20 of the agents (87\%) did not implement any wake action, 2 (9\%) used wake word, and 1 (4\%) got triggered only after playing a beep. 3 papers (13\%) described agents with the ability to recognize user's emotions. 

Three of the surveyed papers sought to understand how children with autism behave with the conversational agents to promote social and communication skills.
\citet{soleiman2014roboparrot} employed a parrot-shaped robot called RoboParrot 
that interacted both with typically-developed children and children with ASD to stimulate to socially interact with the agent during unstructured sessions. Both groups showed interest in the robot, but the autistic children hesitated to approach it and needed to be encouraged by their parents. They observed that when children were accompanied by their parents, they showed more interaction with the robot, but still with limited behavioral repertoire and varieties. 
Similarly, \citet{Cervera2019-un} employed a Nao robot for a 10-months long intervention 
to improve the interest and motivation of children with ASD during activities. 5 children were assigned to the human condition (control group), and 9 of them to the robot condition (intervention group). 
Results showed that there was not significant difference between the two groups, however the intervention group showed positive attitude, such as gaze at the robot more than the control group.
Again, \citet{Chung2021-fl} used a Nao robot 
and ran a task-based training with the technology based on the ABA design protocol and involving 15 children with ASD over 12 weeks. Their results showed that the robot was effective in enhancing the eye contact and verbal initiation of the children with the human caregiver.

Three works also focused on the role of the agent as teacher to learn communication skills from via games.
\citet{So2019-vm} employed a Nao robot 
that they compared with humans in teaching gestural recognition and production to children with ASD. They randomly assigned 23 children with autism to either a robot or a human group. Findings showed that, in both groups, children with ASD accurately recognized and produced the gestures required.
Similarly, \citet{Zhang2019-ma} employed a Nao robot 
to explore how children with and without autism learn social rules through distrust and deception games with the robot. Their study involved 20 children with ASD and 20 neurotypical children. Results showed that children with ASD took longer to learn to distrust and deceive the socially assistive robot than children without ASD, and children with ASD that perceived the robot more human-like had more difficulty in distrusting it. 
Again, \citet{desideri2018using} used a Nao robot 
to explore whether it could enhance engagement and learning achievement running a pilot study with two children with ASD. Their study followed an ABA single-case design, including an intervention period with the socially assistive robot. Their results showed that the robot increased the achievement of the learning goals of both children, however only in one case it also enhanced engagement. 

Two studies focused on the therapeutic intervention, specifically on how an agent can be introduced as a tool to promote the training of social and communication skills.
\citet{pop2013social} employed an animal-shaped socially assistive robot called Probo
and compared its effectiveness with a computer for training social skills of 20 children with ASD. 
The children were randomly allocated to either control group, computer assisted therapy, and robot assisted therapy. 
Findings indicated that using the socially assistive robot to implement social story intervention was more effective for improving the independence of expressing social abilities for the participants.
Similarly, \citet{spitale2020whom} employed both an embodied conversational agent and a socially assistive robot with an sheep shape 
and compared the two agents with humans in a within-subject study in the context of a speech-language therapeutic intervention. They involved 14 typically developed children and 3 children with language disorders. 
Their results showed that the physicality of the agent affected the linguistic performance of children. In addition, children showed a preference towards the physical agent and perceived it as smarted than the virtual one.

Two studies highlighted the potential of using social cues to adapt the agent behavior to the specific user needs.
\citet{tanaka2017embodied} designed a woman embodied conversational agent 
to explore the use of the agent for training social skills and providing feedback to users regarding audiovisual features of their speech, pitch, language, smiling ratio, and yaw. They recruited 18 people from the general population and 10 people with autism spectrum disorders. 
Results indicated the individual improvements of the pre- and post- social skills by both groups of participants suggesting also the effectiveness of real-time audiovisual features extraction, analysis, and exploitation during the interaction. 
Similarly, \citet{khosla2015service} used a Lucy humanoid robot to explore its use by a person with autism during a free-interaction longitudinal trial in a home-based environment. Results establishes that the robot could maintain the engagement high and develop reciprocal relationship with the user through both service personalization and its human-like characteristics involving voice, gestures, and emotion recognition and expression.

Three studies explored conversational agents to promote verbal interaction.
\citet{shimaya2016advantages} used a humanoid robot named CommU 
that freely interacted with  three teenagers with ASD for around 40 minutes for 2-4 days. Participants showed some positive tendencies such as non-echolalic responses and talk about problems related to human relationships while interacting with the agent. Moreover, a quantitative analysis of the utterances in the trials suggested that reduced utterances from a caregiver might positively influence individuals with ASD toward sharing more information.
Similarly, \citet{Shimaya2019-xg} employed the same humanoid socially assistive robot 
to provide individuals with ASD an environment where they could disclose their thoughts and concerns. They ran one pilot study including 2 teens with ASD. Results showed that the participants were able to disclose more during the session with the robot than with the human concealer. 
Again, \citet{smith2021smart} employed Alexa and Google Home 
with 21 individuals with intellectual disabilities and let them free to interact with the technology for about 12 weeks to assess intelligibility gains. Meanwhile, a control group of 22 people did not use any conversational technology. 
Findings showed that the group that received the conversational agents made significantly larger intelligibility gains than the control group.

Finally, we noticed that all but one papers addressing skills on joint attention, emotional sphere, touch confidence, and sensory perception reported that they also addressed communication and social skills in a broader sense.
\citet{axelsson2019participatory} reported that the agent succeeded teaching sign language to participants. \citet{So2020-oq} showed that participants significantly improved their engagement during games and social behavior in daily life. \citet{Huijnen2021-nx} found out that participants significantly increased non-verbal imitation capabilities while interacting with the agent.
\citet{amanatiadis2017interactive} showed that participants improved their social skills during the interaction and accepted the agent as a companion.
Also \citet{Van_Otterdijk2020-ww} proved that participants increased the engagement in social activities after the interaction with the agent. Authors suggested that those results were also due to the personalization of the activities to meet the special needs of each user.
\citet{fachantidis2020tauhe} indicated that during the interaction with the agent there were more incidences of eye contact, proximity and verbal interaction than during the interaction with the human. Additional behaviors (e.g., reduced fidgeting and increased attention and ability to follow instructions) improved during interaction with the conversational technology.
\citet{catania2021toward} showed that the conversational agent had the potential for being used as a stimulus for oral communication and socialization in many activities with participants from low to high severity. Participants’ cognitive, sensory, and language impairments made it difficult for them to autonomously use the agent (e.g., they could not quickly adapt to the schematic communicative protocol involving the use of the wake word) and to make themselves understood by the automatic speech recognition system. Still, authors learned that conversational agents can be considered a valid tool to support the therapy of children with NDD, but to be usable they would need to include customization options and multiple modalities of interaction (e.g., visual and/or touch interface).


To sum up, \textit{results showed that even when a conversational agent is initially designed for a specific therapeutic goal, its use contributes to promote abilities related to communication and social skills in a broader sense. Conversational agents are effective in facilitating the use of gestures, eye contact, verbal initiation, and appropriate social behavior rules. They can also enhance the level of participants’ attention and  engagement during the therapy, produce some positive tendencies, such as non-echolalic responses, and stimulate talks about problems related to human relationships.
Additional findings comprise the following: i) multi-modal interaction, possibility of personalization and customization of some interaction features, human-like characteristics of the agent representation (e.g., voice, shape), and (in social robot) physicality are paramount design features; ii) for some users with NDD is particularly hard to formulate a word or a sentence properly; this affects the agent capability of understanding natural language expressions and react correctly, which in turn makes it difficult for some users to use the agent autonomously; still such inability of the agent is not perceived so frustrating by persons with NDD nor it totally prevent them to engage with the agent; iii) technological features of extraction and analysis of real-time audiovisual features increases the capability of the agent in both the interpretation of human input and the production of output, particularly when user’s input has flaws, e.g., it is incomplete, structurally wrong,  or badly pronounced.}

\subsection{Open research questions (R21)}
In this subsection we describe the follow-up research plans reported of the surveyed papers that, in some cases, remedy some of their studies' limitations, while, in other cases, address some new research questions they opened by advancing the state of the art. 

\citet{Cao2019-yg} shared that they wanted to investigate how conversational technology might become more generally accepted in therapeutic settings.
\citet{desideri2018using} recommended stronger cooperation between the technological and clinical actors of research to further understand the effects of technology-based therapy on the development of cognitive and social skills in people with NDD. 
Indeed, a more active collaboration might generate more effective, targeted, and structured technology-based activities as demanded by \citet{khosla2015service} and \citet{Van_Otterdijk2020-ww}. 
To facilitate such a collaboration, \citet{shimaya2016advantages} wanted to create a framework accessible to non-technical people that aids the creation of conversational agents specifically for people with NDD by making them customizable and adaptable to each user's needs through a modular architecture.

The research plan includes also an extensive study of interaction design to understand how different features influence conversational technology's usability, engagement, and therapeutic efficacy.
\citet{tanaka2017embodied} suggested exploring the impact of changing the gender of the agent in users' perception and performance in the activity.
\citet{spitale2020whom} wanted to understand further which is the best embodiment for a conversational agent for NDD.
\citet{catania2021toward} aimed to figure out which wake action is best depending on the user and the scenario.
\citet{desideri2018using} wanted to clarify in which role (e.g., reinforcer vs. mediator) a conversational agent could be more effective in improving the intervention.
Also, a deeper understanding is needed among preferences of neurotypical people and people with NDD toward different additional input and output interaction modalities and their combinations \cite{axelsson2019participatory, soleiman2014roboparrot, catania2021toward}. Specific to output interaction modes, \citet{axelsson2019participatory} stressed the need to improve the expression skills of conversational agents (e.g., through the face) since this might improve the user experience at large.

Current conversational agents for the therapy of people with NDD are far from being autonomous. Improving automatic speech recognition modules would enable more self-ruling conversational agents, which would reduce the human task load.
For this reason, future work should pursue improving certain technical drawbacks, especially regarding speech recognition problems, also considering more severe cases of speech impairments \cite{amanatiadis2017interactive, tanaka2017embodied, smith2021smart, soleiman2014roboparrot, fachantidis2020tauhe}. As a possible research direction, \citet{smith2021smart} suggested implementing a \textit{voice profile} option, which allows the speech recognition model to adapt to individual voices.
For those conversational agents where the real-time remote control is not provided yet, \citet{axelsson2019participatory} highlighted the need to offer this functionality in the future to empower therapists with more control during interventions.

A shared feeling among the researchers is the lack of stability of the results they obtained and the consequent need for more robust findings. For this reason, some authors reported that they would like to focus on more extended empirical studies in the future \cite{smith2021smart, catania2021toward, Huijnen2021-nx, desideri2018using, Cervera2019-un, So2020-oq, fachantidis2020tauhe, Zhang2019-ma} and in more ecological settings out of the lab \cite{So2020-oq}.
In this way, the novelty effect of the technology would be mitigated, and researchers could finally assess if the social skills learned by participants persist in time and in different social contexts with people \cite{Cervera2019-un, desideri2018using, tanaka2017embodied, Chung2021-fl, fachantidis2020tauhe}.
Authors also declared they want to recruit more people for future studies \cite{pop2013social, catania2021toward, Huijnen2021-nx, Cervera2019-un, marino2020outcomes, So2020-oq, Zhang2019-ma}. It is essential to include people of all ages \cite{So2020-oq} and with a wide range of disturbs and special needs \cite{tanaka2017embodied}. Participants should also be balanced in gender \cite{Chung2021-fl, marino2020outcomes}. So far, most of the people involved in the analyzed studies were male and consequently findings did not provide a precise picture of the benefits and fatigue of males and females while interacting with conversational technologies. Finally, \citet{Zhang2019-ma} and \citet{So2020-oq} wanted to involve people from different cultures to assess if results change depending on linguistic, cultural, or social variables.

To sum up, \textit{results showed that researchers in the field plan to address a number of goals in the future: exploring different human-computer interaction aspects, improving the technology to achieve autonomous systems, obtaining more reliable empirical results, and facilitating the collaboration among people with a technical background and NDD experts.}

\section{Recommendations}
\label{discussion}
Starting from the lessons we learned from this literature review, we drew some recommendations that may be helpful to interaction designers, developers, and researchers approaching the field of conversational technology for supporting the therapy of people with NDD. 
Recommendations concern the \textit{design} of conversational agents and their empirical evaluation \textit{method}.


\subsection{Design}

From a design perspective, we paid strong attention on some relevant design features of the conversational agents in the surveyed papers, such as the embodiment, shape, gender, interaction modalities, wake action, and emotion recognition skills.

\begin{enumerate}
    \item Our findings showed that most of the surveyed papers, such as \cite{pop2013social, khosla2015service, soleiman2014roboparrot, desideri2018using, Chung2021-fl}, adopted a socially assistive robot (i.e., physical embodiment) for training or assessing skills of the users. Also, the surveyed papers explored embodied conversational agents, and intelligent personal assistants. 
    In general, all of them proved to have good potential in supporting the activities with the users with NDD, but it is still unclear which one is best for usability and therapeutic effectiveness. 
    In this regard, a previous work \cite{spitale2020multicriteria} proposed a method for comparing the embodiment of conversational agents for NDD in a systematic way without any experimental trial. As a case study, they compared socially assistive robots, embodied conversational agents, and intelligent personal assistants for one-to-one therapy of children with autism and found out that embodied conversational agents are the best solution since they satisfy a reasonable set of achievements (e.g., maximize safety, maximize scalability, maximize configurability). The validity of this result is still to be empirically verified. 
    
    \textit{We recommend designers to choose the embodiment for the conversational agent as one of the first steps of the design process since it has significant design implications.
    For example, each embodiment enables different interaction modalities both from and for the user (e.g., socially assistive robots can offer physicality-based interactions that embodied conversational agents cannot). We recommend also considering both the skills of the users and the goals to be achieved with the technology in order to choose the best embodiment depending on the usage scenario. For instance, IPAs are characterized by playing the role of users' assistants and are not suitable in those scenarios requiring a companion, a moderator, or trainer. Embodied conversational agents can generally offer a controlled interaction space that helps the user to maintain a high level of attention during task-based activities. We do not recommend using disembodied conversational agents, which have never been explored so far due to their abstract nature that, according to the shared opinion of NDD experts, cannot maintain a high level of user attention and engagement during the activities.}

    \item We found out that most of the surveyed papers (e.g., \cite{So2019-vm, axelsson2019participatory, Cao2019-yg}) chose a bio-inspired shape, specifically with a human-like appearance. From literature, we know that the shape of a conversational agent can affect its perception by users \cite{shape}, and also their performance in the proposed activities with the technology \cite{ricks2010trends}. 
    Although most of the conversational agents in the review had a human-like appearance, there is no clear consensus on what a conversational agent for NDD therapy should look like.  
    \cite{ricks2010trends} reports that a humanoid shape may be preferred when the goal of the activities with the conversational agent is the generalization of training behaviors, given that similarity to humans facilitates this aim. Still, simplistic shapes (e.g., cartoon-like embodied conversational agents) may be used because they successfully hold the attention of people with NDD. Non-humanoid shapes fit the needs of people feeling safer in an environment with less human-like stimuli. 
    
    \textit{We recommend for a long-term study to begin with the use of a simple-shaped conversational agent to get the user involved in the activity and, after some sessions, replace it with a more realistic agent (e.g., human-like).}

    \item Our findings showed that the surveyed papers almost equally adopted neutral, male, or female genders. 
    Since conversational agents are generally attributed to human-like characteristics, they are often assumed to have a gender made explicit with the appearance or the voice. 
    To the best of our knowledge, there is no evidence on how the gender of the conversational agent impacts the usability or performance of subjects with NDD. However, from literature, we know that generally, gender sets up expectations impacting the user experience, for example, influencing the level of comfort, fun, and agent's intelligence perceived by users \cite{niculescu2010agent}. 
    Some neuropsychologists advocated for conversational agents to have a neutral robotic voice to prevent users from confusing them with humans and becoming reluctant to interact \cite{axelsson2019participatory}. 
    
    \textit{We recommend employing a gender-neutral agent as it may help users with NDD to become more aware that they are interacting with a machine and thus reduce their social fear and possible sense of uncanny valley \cite{skjuve2019help}.}
    \item We found out that the surveyed papers were able to provide information to the user with speech, gestures, gaze, walking, playing some sounds, performing facial expressions, displaying content on a screen, switching on and off some lights, and touching. 
    They captured information from the user from speech, gestures, gaze, touch, facial expressions, biosensors, and movements in space. 
    The target population has different disorders and a wide range of special needs, and the scenarios for using a conversational agent during all possible therapeutic activities are many. 
    To the best of our knowledge, there is no evidence defining which interaction modalities are best for people with NDD in general. 
    The surveyed papers suggested that conversational agents have a potential in the context of the therapy of people with NDD also because they are multi-modal by nature \cite{google} and thus compensate the user's difficulties in a single modality by offering other ways to interact. 
    
    \textit{We recommend enhancing the conversational agent with as many interaction modalities as possible and implementing the system in a modular way so that each modality can be enabled or disabled on point depending on the user's needs and the different scenarios without impacting the functioning of the system.}

    \item Our findings showed that the majority of agents described in the surveyed papers (e.g., \cite{mower2011rachel, pop2013social, bekele2016multimodal}) did not employ any wake action, and they designed the conversations with the user as natural human-human interactions. This solution is certainly the most intuitive and natural for the user, but it requires the agent to be smart enough to handle turn-taking. 
    Other authors, such as \citet{catania2021toward}, used a wake word to initiate the speech-based interaction for children with NDD. Their findings showed that some of those children did not find trivial to interact with the agent by combining the wake action and their spoken response since these parts are not semantically connected. 
    A previous study in the literature \cite{catania2020best} proposed to use identical actions both to wake up and to put to sleep the conversational agent, and provided a theoretical argument based on the theory of \textit{partner-perceived communication} \cite{costantino,light2007aac}, which states that the predictability and repetitiveness make it possible to better give meaning to the sentences even for those people with complex communication needs. 
    
    \textit{We recommend considering the advantages and disadvantages of each wake action based on users' special needs, the goals of the activities with the technology, and the context in which it will be used.
    Interaction designers might consider using different wake actions (e.g., vocal, tactile, visual, event-based, or motion-based) based on the user scenario.
    We recommend also developers to empower conversational agents with advanced features such as real-time message-filtering and turn-taking capabilities to enable interaction designers to remove wake actions in automated systems and achieve a more natural interaction.} 

    \item We found out that only one of the surveyed papers \cite{khosla2015service} endowed its conversational agent with emotional capabilities and just few of the surveyed papers \cite{tanaka2017embodied, bekele2016multimodal} extracted users' social cues from video and audio recordings to adapt the interaction with the users and showed the effectiveness of this practice. Features were about the user's speech, pitch, language, facial expression, smiling ratio, and yaw.
    
    In contrast with the majority of the papers, \textit{we recommend designers to empower the agents with the capability to extract users' social cues. In this way, the agents could understand them better and adapt the flow of the conversation based on both verbal and nonverbal information. 
    We recommend also combining different features and interpreting them to detect social behaviors (e.g., echolalic responses, emotions, attention level).} In this way, NDD experts could guide designers on how they want the conversational agent to behave in different situations in response to any of these well-known behaviors. Last but not least, understanding user emotions from different channels may enable new scenarios such as the development of conversational agents that work as emotional trainers to further enhance users' recognition and expression skills.
\end{enumerate}

\subsection{Method}
From a methodological perspective, this literature review onset guidelines and recommendations on the demographics to recruit and other study methodology choices to be made for running a well-designed empirical research on conversational agents for the therapy of people with NDD.

\begin{enumerate}
    \item 
    Our findings showed that many of the surveyed studies (e.g., \cite{mower2011rachel, soleiman2014roboparrot, khosla2015service}) have only included small sample size of people with NDD. We understood that participants with NDD are not easy to recruit in general and sometimes the skepticism of their caregivers in adopting innovative technologies during the therapy sessions impact their participation to research studies \cite{kaburlasos2019social}. 
    
    For this reason, \textit{we recommend researchers to involve caregivers, parents, and therapists during every step of the research, from study design to data analysis. In this way, caregivers might see the potential of the technology and provide helpful feedback on the prototypes even when they are still in progress.}

    \item We observed that the majority of the surveyed papers \cite{bekele2016multimodal, tanaka2017embodied, Cao2019-yg} have largely explored the use of conversational agents with a male population with NDD, and the gender distribution across the study is highly unbalanced. This result suggested that researchers could recruit male participants easier than female ones. The main reason is that females and males significantly differ in their probability to develop neurodevelopmental disorders \cite{pinares2018sex}. In fact, males are more likely to onset neurodevelopmental disorder than females. 
    
    To achieve a gender distribution balance and have a deeper and broader understanding of conversational technology interacting with people with NDD, \textit{we recommend considering that it is more difficult to recruit females than males for an empirical study. }

    \item Our findings suggested that most of the surveyed studies, such as \cite{Zhang2019-ma, spitale2020whom, pop2013social}, have explored the potential of conversational agents for children (less than 18 years old), while only few studies investigated the impact of those technologies for adults. 
    We know that early therapy has a stronger impact than late therapy and when people get early interventions they may require less or no assistance as they get older \cite{powers1992early}. For this reason, we understand that researchers prioritized research for youngsters. 
    
    Still, \textit{we recommend researchers to run some empirical studies that involve adults as well to meet their specific needs and to have a more comprehensive understanding of NDD in a broader sense.}

    \item We observed that the majority of the study surveyed (e.g., \cite{Van_Otterdijk2020-ww,Zhang2019-ma, soleiman2014roboparrot, khosla2015service}) only focused on autism among all neurodevelopmental disorders. We understand that researchers obtained promising results for this population and want to explore it further. 
    
    However, \textit{we recommend the HCI community to investigate also on the other disorders (e.g., intellectual disability, developmental language disorders), to provide a more comprehensive understanding of the potential of conversational technologies for people with NDD.}
\end{enumerate}

Regarding the empirical study evaluation, we recommend the following.
    
\begin{enumerate}
    \item The results of the surveyed papers showed that the main skills of people with NDD that were addressed so far regard communication and socialization. In some cases, conversational agents showed that they can promote social interaction enhancing eye contact, proximity, and verbal interaction more than during human-human interaction \cite{Huijnen2021-nx, fachantidis2020tauhe}. However, other studies \cite{desideri2018using, axelsson2019participatory, So2020-oq} revealed that 
    people with NDD did not significantly improved specific skills (e.g., \cite{Cervera2019-un}) during the interaction with an agent - compared with the interaction with a human. 
    
    Still, \textit{we recommend researchers to employ conversational agents for both free-interaction and task-based activities since they appear to produce same or higher engagement than humans and may promote users's social and communication skills despite these were not originally supposed to be addressed within the activity (e.g., turn-taking).}

    \item Our literature review showed that a broad variety of design study approaches have been used by researchers, from within-subject studies to exploratory ones. 
    %
    %
    %
    
    \textit{We recommend to design empirical studies with a randomized control trial (RCT) approach to avoid biases, and to run longitudinal studies to consider the long-term impacts of the therapy with the technology.}

    \item 
    Our review suggested that in recent years researchers (e.g., \cite{Chung2021-fl, smith2021smart, catania2021toward}) tend to use more autonomous conversational agents, probably because of the progress of natural language processing technology. Still many authors adopted Wizard-of-Oz prototypes to evaluate their research questions.
    
    \textit{We recommend to drive future research on autonomous agents because their evaluation enables a better understanding of the real capabilities of the agents during the interaction with people with NDD.} Still, we acknowledge that the Wizard-of-Oz approach can be a useful evaluation method for exploring design aspects that are independent from the technological capabilities (e.g., the agent's shape). 

    \item 
    This literature review showed that researchers adopted very different data collection approaches, often combining qualitative and quantitative analysis. For example, \citet{Huijnen2021-nx} found out that their mixed-approach of analysis that combines qualitative and quantitative methods was really effective to confirm and strength their study results. Also, we learned the importance of collecting subjective data from caregivers' perspectives.
    
    \textit{We recommend to design empirical studies with a combination of qualitative and quantitative methods to gain multiple perspectives to evaluate the efficacy of conversational technologies to improve the skills of people with NDD. }
\end{enumerate}

\subsection{Checklist for systematically reporting of empirical studies}

We provide a non-exhaustive checklist for HCI researchers who want to run an empirical study with a conversational agent to support the therapy of people with NDD (see Table \ref{tab:checklist}). We recommend researchers to use this checklist to ensure a high-quality report in their papers without missing the most relevant information about the design features of the agent and the methodology aspects of the study, since we noticed from the papers of our review that sometimes it happens.

\begin{table}[h!]
    \centering
    \begin{tabular}{|l|l|l|}
     \hline
    \textbf{Category}     & \textbf{List of items} & \textbf{Example}\\
    \hline
    \textbf{Conversational Agents}
        &$\square$ Agent's goal & assessment, training \\
        &$\square$ Embodiment & physical, virtual, disembodied etc. \\
        &$\square$ Shape of the agent & humanoid, animal-like, vegetables, etc. \\
        &$\square$ Gender of the agent & female, male, neutral etc. \\
        &$\square$ Input interaction modalities & speech-based, face-based, gesture-based etc. \\
        &$\square$ Output interaction modalities & speech-based, face-based, gesture-based etc. \\
        &$\square$ Wake action & none, wake word, buzzer etc. \\    
        &$\square$ Additional capabilities & emotion recognition etc. \\
         \hline
    \textbf{Demographics} &$\square$ Sample size &  \\
    &$\square$ Age & pre-schooler, grade-scholer, teens, adults etc. \\
    &$\square$ Gender distribution & equal etc. \\
    &$\square$ Diagnosis & ASD, ID, DLD etc. \\
    &$\square$ Prior experience with agents & zero, a little, expert, etc. \\
         \hline
    \textbf{Empirical Study} &$\square$ Setting & home, therapeutic center etc. \\
        &$\square$ Special needs addressed & communication, social, emotional, etc. \\
        &$\square$ Study design & within-subject, between-subject, RCT etc. \\
        &$\square$ Study duration & days, weeks etc. \\
        &$\square$ Nature of the prototype & wizard-of-Oz, automated, semi-automated etc. \\
        &$\square$ Interaction type & task-based, free interaction \\
        &$\square$ Collected data type & objective, subjective etc. \\
        &$\square$ Data collection method & interview, questionnaires, etc. \\
         \hline
    \textbf{Data analysis} &$\square$ Detailed procedure &  \\
    &$\square$ Analytic tool & python library, statistical program, etc. \\
    &$\square$ Link to script & \underline{\url{https:\\\\my-custom-script.com}}\\
         \hline
    
    \end{tabular}
    \caption{Checklist to systematically report empirical studies on conversational agents for people with NDD.}
    \label{tab:checklist}
\end{table}

\section{Research agenda}
We propose a research agenda on conversational agents for the therapy of people with NDD based on the current state of the art and open research challenges identified in the review.

\begin{enumerate}
    \item Future research will need to investigate the democratization of conversational agents' development. We mean that conversational technology development must be made easily and widely accessible without requiring conversational agent developers to have in-depth software engineering knowledge. Conversational agents' development democratization is fundamental since conversational agents are not emerging as mere conventional technologies but as agents operating in social contexts. So, in the contest of conversational agents for NDD, they should be created by psychologists, therapists, and interaction designers who are experts in the NDD and HCI areas and can better address the special needs of each individual and the goals of the different therapeutic activities. 
    
    \item It urges a deeper understanding of the preferences of people with NDD toward different design features of conversational agents. Indeed, we know about some general design guidelines for conversational agents, but universally recognized guidelines specific to NDD are still lacking. Features to investigate include different input and output interaction modalities, wake actions to trigger the system,  gender, embodiment, and shape of the agent. Also, we observed that conversational agents for NDD generally lack the ability to recognize users' emotions. Given that this ability could be helpful in generating responses that are more consistent with the context of the conversation, we believe that future research should focus on studying emotional recognition from the speech and face of users with NDD. Conversational agents could integrate this technology, adapt the dialogues in real-time based on emotion analysis results, and obtain emotional feedback on the user experience. 
    Improving the user experience means improving conversational agents' usability, engagement, and therapeutic effectiveness for NDD therapy and thus establishing them as a valid solution in this domain.
    
    \item Future research should address some technological weaknesses, including voice recognition issues considering more severe speech impairments cases. 
    While the increased computational power of deep learning systems and the availability of large training datasets has improved the accuracy of ASR systems, their performance is still insufficient for many people with speech disorders, rendering the technology unusable for many of the speakers who could benefit the most. Also, continuous misunderstandings are likely to lead them to frustration, making the use of conversational agents ineffective in therapeutic settings.
    It urges to implement personalized ASR models for recognizing a wide range of speech impairments and severities, with potential for making ASR available to a wider population of users.
\end{enumerate}

\section{Limitations}
This study has some limitations. 

First, the focus of our research was narrow and centered on conversational agents for training and assessing the communication and social skills of people with NDD. We used strict criteria regarding what research designs to include (i.e., we considered empirical studies only), potentially excluding valuable findings from non-experimental works. Any possible interest in this research can only be inferred from future research directions taken by the HCI community, but the trend of papers over the years that we highlighted in this review is promising.

Although we followed a systematic and standardized method of selection and analysis, some relevant papers may have been unintentionally excluded from our review.
We did not use all possible databases, and the search query may not have returned all relevant records. We manually added papers as needed, but just follow-up work can confirm the validity of our selections.

Finally, because of the small number of participants for each selected study and their heterogeneity in terms of the disorders they addressed, this review's results might be considered provisional, waiting for future more extensive studies on the subject. 
Still, we were able to draw up high-level design and methodological recommendations from the lessons we learned from the surveyed studies and we believe that the knowledge we report may be helpful to new researchers approaching this domain in the future.

\section{Conclusion}
\label{conclusion}
The adoption of conversational agents for people with NDD is emerging as promising to assess and stimulate their skills in therapeutic interventions. This survey of a number of relevant publications in the last decade highlighted i) the large amount and variety of therapeutic goals addressed by the conversational agents used in the therapy of people with NDD – which is not surprising, considering the broad variety of disorders and impairments associated to NDD, and ii) the wide gamut of different methodological approaches adopted in their  design and evaluation.

Still, our analysis also provides a view at large of some general solutions in the current state of the art. From a design perspective, results showed, for example, that most of the surveyed papers focused on male children with autism to train their communication and social skills. Half of those papers adopted both speech and gestures as either input and output interaction modalities and they designed the conversation as a natural human-human interaction without employing any wake actions. In addition, most of the surveyed papers adopted humanoid socially assistive robots and they were not endowed with emotional capabilities. Regarding the methods adopted in empirical studies, our findings showed that a task-based therapeutic approach with autonomous agents is adopted in most cases. Also, the majority of studies were short-term (less than 9 sessions) and single-subject, and the main sources of data collected were qualitative (textual notes, video and audio recordings) and were then manually annotated for coding and further analysis.

From our analysis we extracted the main lessons learned and elaborated a set of methodological recommendations concerning both the design of conversational agents for NDD therapy and their empirical evaluation. Our review also highlighted the gaps left unsolved in the state of the art, which we regard as opportunities for future investigations. 

\bibliographystyle{ACM-Reference-Format}
\bibliography{01_bib_the_real}


\newpage
\appendix
\section{The surveyed papers at a glance}
\label{appendix}
\begin{table}[ht!]
\centering
\small
\captionsetup{width=\textwidth}
\caption{Agent's features of the surveyed papers in terms of embodiment, gender, shape, and emotion recognition capabilities. If a row has not checks it means that we did not retrieve any information about that specific variable in the corresponding paper. \\
\textit{Legend}: socially assistive robot (SAR), embodied conversational agent (ECA), intelligent personal assistant (IPA), male (M), female (F), neutral (N).}
\label{tab:agent}
\begin{sideways}
\begin{tabular}{l|lll|lll|llll|l}
\hline
\multicolumn{1}{l}{{\ul \textbf{Ref}}} & \multicolumn{3}{c}{{\ul \textbf{Agent's}}} & \multicolumn{3}{c}{{\ul \textbf{Agent's}}} & \multicolumn{4}{c}{{\ul \textbf{Agent's}}}                                                      & \multicolumn{1}{c}{{\ul \textbf{Emotional}}} \\

\multicolumn{1}{c}{} & \multicolumn{3}{c}{\ul \textbf{Embodiment}}  & \multicolumn{3}{c}{{\ul \textbf{ Gender}}} & \multicolumn{4}{c}{{\ul \textbf{Shape}}}                                                      & \multicolumn{1}{c}{{\ul \textbf{recognition}}}\\
\hline
                   & \textbf{SAR}  & \textbf{ECA}  & \textbf{IPA} & \textbf{M}       & \textbf{F}       & \textbf{N}      & \textbf{Humanoid} & \textbf{Animal-} & \textbf{Vegetable} & \textbf{Speaker} & \textbf{Facial }              \\
                                      &   &  &  &     &      &     & & \textbf{like} &  &  & \textbf{expression}              \\
                   \hline
\citet{mower2011rachel}, \citeyear{mower2011rachel}             &               & x             &              &                  & x                &                                      & x                 &                      &                  &                    &                                                           \\
\citet{pop2013social}, \citeyear{pop2013social}                & x             &               &              &                  &                  &                                    &                   & x                    &                  &                    &                                                            \\
\citet{soleiman2014roboparrot}, \citeyear{soleiman2014roboparrot}                & x             &               &              &                  &                  & x                               &                   & x                    &                  &                    &                                                            \\
\citet{khosla2015service}, \citeyear{khosla2015service}                & x             &               &              &                  &                  &                                  & x                 &                      &                  &                    & x                                                       \\
\citet{shimaya2016advantages}, \citeyear{shimaya2016advantages}                & x             &               &              & x                &                  &                                     & x                 &                      &                  &                    &                                                   \\
\citet{bekele2016multimodal}, \citeyear{bekele2016multimodal}           &               & x             &              & x                & x                &                                   & x                 &                      &                  &                    &                                                        \\
\citet{amanatiadis2017interactive}, \citeyear{amanatiadis2017interactive}                & x             &               &              &                  &                  &                                     & x                 &                      &                  &                    &                                                      \\
\citet{tanaka2017embodied}, \citeyear{tanaka2017embodied}               &               & x             &              &                  & x                &                                   & x                 &                      &                  &                    &                                                           \\
\citet{desideri2018using}, \citeyear{desideri2018using}                & x             &               &              &                  &                  &                                  & x                 &                      &                  &                    &                                                         \\
\citet{axelsson2019participatory}, \citeyear{axelsson2019participatory}              & x             &               &              &                  &                  & x                               & x                 &                      &                  &                    &                                                        \\
\citet{Cao2019-yg}, \citeyear{Cao2019-yg}                & x             &               &              &                  &                  &                                  & x                 &                      &                  &                    &                                                   \\
\citet{So2019-vm}, \citeyear{So2019-vm}                & x             &               &              &                  & x                &                                   & x                 &                      &                  &                    &                                                        \\
\citet{Shimaya2019-xg}, \citeyear{Shimaya2019-xg}                & x             &               &              & x                &                  &                                & x                 &                      &                  &                    &                                                          \\
\citet{Zhang2019-ma}, \citeyear{Zhang2019-ma}                & x             &               &              &                  &                  & x                             & x                 &                      &                  &                    &                                                          \\
\citet{Cervera2019-un}, \citeyear{Cervera2019-un}                    & x             &               &              & x                &                  &                                   & x                 &                      &                  &                    &                                                      \\
\citet{So2020-oq}, \citeyear{So2020-oq}                & x             &               &              &                  &                  &                        & x                 &                      &                  &                    &                                                            \\
\citet{fachantidis2020tauhe}, \citeyear{fachantidis2020tauhe}                & x             &               &              &                  &                  &                               &                   &                      &                  & x                  &                                     \\
\citet{marino2020outcomes}, \citeyear{marino2020outcomes}                   & x             &               &              &                  &                  &                            x&                  &                      &                  &                    &                                                         \\
\citet{Chung2021-fl}, \citeyear{Chung2021-fl}                & x             &               &              &                  &                  &                                 & x                 &                      &                  &                    &                                                             \\
\citet{Van_Otterdijk2020-ww}, \citeyear{Van_Otterdijk2020-ww}                & x             &               &              &                  &                  &                                  & x                 &                      &                  &                    &                                                       \\
\citet{spitale2020whom}, \citeyear{spitale2020whom}                 & x             & x             &              &                  &                  & x                           &                   & x                    &                  &                    &                                                             \\
\citet{Huijnen2021-nx}, \citeyear{Huijnen2021-nx}                & x             &               &              &                  &                  &                                  & x                 &                      &                  &                    &                                                          \\
\citet{smith2021smart}, \citeyear{smith2021smart}                &               &               & x            &                  &                  &                                 &                   &                      &                 &   x                 &                                                          \\
\citet{catania2021toward}, \citeyear{catania2021toward}                &               &               & x            & x                &                  &                               &                   &                      &                 &   x                 &                                                           \\
\hline
\textbf{Count}     & 19            & 4             & 2            & 5                & 4                & 4                                 & 18                & 3                    & 2                & 1                  & 1                                   \\
\hline
\end{tabular}
\end{sideways}
\end{table}

%

\begin{table}[h!]
\centering
\small
\caption{Agent's feature of the surveyed papers in terms of wake action and the nature of the prototype automation.  If a row has not checks it means that we did not retrieve any information about that specific variable in the corresponding paper. [continued]}
\label{tab:agent2}
\begin{sideways}
\begin{tabular}{l|lll|lll}
\hline
\multicolumn{1}{l}{{\ul \textbf{Ref}}} & \multicolumn{3}{c}{{\ul \textbf{Wake action}}}                           & \multicolumn{3}{c}{{\ul \textbf{Nature of the prototype}}}               \\
\hline
                   & \textbf{Buzzer} & \textbf{None} & \textbf{Wake word} & \textbf{Wizard-of-Oz} & \textbf{Semi-automated} & \textbf{Automated} \\
                   \hline
\citet{mower2011rachel}, \citeyear{mower2011rachel}             &                 & x                                  &                    & x                     &                          &                    \\
\citet{pop2013social}, \citeyear{pop2013social}                &                 & x                                  &                    & x                     &                          &                    \\
\citet{soleiman2014roboparrot}, \citeyear{soleiman2014roboparrot}                &                 & x                                  &                    & x                     &                          &                    \\
\citet{khosla2015service}, \citeyear{khosla2015service}                   &                 & x                                  &                    &                       &                          & x                  \\
\citet{shimaya2016advantages}, \citeyear{shimaya2016advantages}                   &                 & x                                  &                    & x                     &                          &                    \\
\citet{bekele2016multimodal}, \citeyear{bekele2016multimodal}                   &                 & x                                  &                    &                       &                          & x                  \\
\citet{amanatiadis2017interactive}, \citeyear{amanatiadis2017interactive}                 & x               &                                    &                    &                       &                          & x                  \\
\citet{tanaka2017embodied}, \citeyear{tanaka2017embodied}                 &                 &                                    &                    &                       &                          & x                  \\
\citet{desideri2018using}, \citeyear{desideri2018using}                       &                 & x                                  &                    &                       & x                        &                    \\
\citet{axelsson2019participatory}, \citeyear{axelsson2019participatory}                      &                 & x                                  &                    & x                     &                          &                    \\
\citet{Cao2019-yg}, \citeyear{Cao2019-yg}                   &                 & x                                  &                    &                       &                          & x                  \\
\citet{So2019-vm}, \citeyear{So2019-vm}                &                 & x                                  &                    &                       &                          & x                  \\
\citet{Shimaya2019-xg}, \citeyear{Shimaya2019-xg}                  &                 &                                    &                    & x                     &                          &                    \\
\citet{Zhang2019-ma}, \citeyear{Zhang2019-ma}                 &                 & x                                  &                    &                       &                          &                    \\
\citet{Cervera2019-un}, \citeyear{Cervera2019-un}                &                 &                                    &                    &                       & x                        &                    \\
\citet{So2020-oq}, \citeyear{So2020-oq}                    &                 & x                                  &                    &                       &                          & x                  \\
\citet{fachantidis2020tauhe}, \citeyear{fachantidis2020tauhe}                    &                 & x                                  &                    & x                     &                          &                    \\
\citet{marino2020outcomes}, \citeyear{marino2020outcomes}                &                 & x                                  &                    & x                     &                          &                    \\
\citet{Chung2021-fl}, \citeyear{Chung2021-fl}                    &                 & x                                  &                    &                       &                          & x                  \\
\citet{Van_Otterdijk2020-ww}, \citeyear{Van_Otterdijk2020-ww}                   &                 & x                                  &                    &                       & x                        &                    \\
\citet{spitale2020whom}, \citeyear{spitale2020whom}                     &                 & x                                  &                    & x                     &                          &                    \\
\citet{Huijnen2021-nx}, \citeyear{Huijnen2021-nx}                  &                 & x                                  &                    &                       & x                        &                    \\
\citet{smith2021smart}, \citeyear{smith2021smart}                      &                 &                                    & x                  &                       &                          & x                  \\
\citet{catania2021toward}, \citeyear{catania2021toward}                   &                 &                                    & x                  &                       &                          & x                  \\
\hline
\textbf{Count}     & 1               & 18                                 & 2                  & 9                     & 4                        & 10        \\
\hline
\end{tabular}
\end{sideways}
\end{table}

\begin{table}[h!]
\centering
\small
\caption{Agent's interaction modalities of the surveyed papers for the input and the output in terms of speech, gesture, movement, facial expression, biological signal, touch, gaze, light, and sound.  If a row has not checks it means that we did not retrieve any information about that specific variable in the corresponding paper.\\
\textit{Legend}: speech (S), gestures and signing (G\&S), movement (M), facial expression (FE), biological signal (BS), touch (T), gaze and head (G\&H); for the output: speech (S), gesture and signing (G\&S), movement (M), facial expression (FE), screen (Sc), touch (T), head and gaze (H\&G), light (L), sound (So). }
\label{tab:interaction}
\begin{sideways}

\begin{tabular}{l|lllllll|lllllllll}
\hline
\multicolumn{1}{l}{{\ul \textbf{Ref}}} & \multicolumn{7}{c}{{\ul \textbf{Input interaction   modality}}}                                                                                                     & \multicolumn{9}{c}{{\ul \textbf{Output interaction   modality}}}                                                                                                                           \\
\hline
                   & \textbf{S} & \textbf{G\&S} & \textbf{M} & \textbf{FE} & \textbf{BS} & \textbf{T} & \textbf{G\&H} & \textbf{S} & \textbf{G\&S} & \textbf{M} & \textbf{FE} & \textbf{Sc} & \textbf{T} & \textbf{H\&G} & \textbf{L} & \textbf{So} \\
                   \hline
\citet{mower2011rachel}, \citeyear{mower2011rachel}              & x               & x                            &                   &                            &                            & x              &                       & x               &                              &                   &                            & x               &                &                       &                &                \\
\citet{pop2013social}, \citeyear{pop2013social}                & x               & x                            &                   &                            &                            &                &                       & x               &                              &                   & x                          &                 &                & x                     &                &                \\
\citet{soleiman2014roboparrot}, \citeyear{soleiman2014roboparrot}                & x               & x                            & x                 &                            &                            &                &                       & x               & x                            &                   &                            &                 & x              &                       &                &                \\
\citet{khosla2015service}, \citeyear{khosla2015service}                   & x               &                              &                   & x                          &                            & x              &                       & x               &                              & x                 &                            & x               &                &                       &                &                \\
\citet{shimaya2016advantages}, \citeyear{shimaya2016advantages}                   & x               &                              &                   &                            &                            &                & x                     & x               &                              &                   &                            &                 &                & x                     &                &                \\
\citet{bekele2016multimodal}, \citeyear{bekele2016multimodal}                   & x               &                              &                   &                            & x                          &                & x                     & x               & x                            &                   & x                          &                 &                & x                     &                &                \\
\citet{amanatiadis2017interactive}, \citeyear{amanatiadis2017interactive}                 & x               & x                            &                   &                            &                            &                &                       & x               & x                            &                   &                            &                 &                &                       &                &                \\
\citet{tanaka2017embodied}, \citeyear{tanaka2017embodied}                 & x               &                              &                   & x                          &                            &                &                       & x               &                              &                   &                            & x               &                &                       &                &                \\
\citet{desideri2018using}, \citeyear{desideri2018using}                       & x               & x                            &                   &                            &                            &                &                       & x               & x                            & x                 &                            &                 &                &                       &                & x              \\
\citet{axelsson2019participatory}, \citeyear{axelsson2019participatory}                      & x               & x                            &                   &                            &                            &                &                       & x               & x                            &                   &                            & x               &                &                       & x              & x              \\
\citet{Cao2019-yg}, \citeyear{Cao2019-yg}                   & x               &                              &                   &                            &                            &                &                       & x               & x                            & x                 &                            &                 &                & x                     &                &                \\
\citet{So2019-vm}, \citeyear{So2019-vm}                & x               & x                            &                   &                            &                            &                &                       & x               & x                            &                   &                            & x               &                &                       &                &                \\
\citet{Shimaya2019-xg}, \citeyear{Shimaya2019-xg}                  & x               &                              &                   &                            &                            &                &                       & x               & x                            &                   &                            &                 &                & x                     &                &                \\
\citet{Zhang2019-ma}, \citeyear{Zhang2019-ma}                 & x               & x                            &                   &                            &                            &                &                       & x               & x                            & x                 &                            &                 &                & x                     &                &                \\
\citet{Cervera2019-un}, \citeyear{Cervera2019-un}                    & x               &                              &                   &                            &                            & x              &                       & x               & x                            &                   & x                          &                 &                &                       &                & x              \\
\citet{So2020-oq}, \citeyear{So2020-oq}                    & x               &                              &                   &                            &                            &                &                       & x               & x                            & x                 &                            &                 &                &                       &                &                \\
\citet{fachantidis2020tauhe}, \citeyear{fachantidis2020tauhe}                    & x               & x                            &                   &                            &                            & x              & x                     & x               &                              &                   & x                          &                 &                & x                     &                &                \\
\citet{marino2020outcomes}, \citeyear{marino2020outcomes}                   & x               &                              &                   &                            &                            &                &                       & x               & x                            & x                 &                            &                 &                & x                     &                &                \\
\citet{Chung2021-fl}, \citeyear{Chung2021-fl}                    & x               & x                            &                   &                            &                            &                & x                     & x               & x                            & x                 &                            &                 &                & x                     &                &                \\
\citet{Van_Otterdijk2020-ww}, \citeyear{Van_Otterdijk2020-ww}                   & x               & x                            &                   &                            &                            &                & x                     & x               & x                            &                   &                            &                 &                &                       &                &                \\
\citet{spitale2020whom}, \citeyear{spitale2020whom}                     & x               &                              &                   &                            &                            &                &                       & x               &                              &                   & x                          &                 &                &                       &                &                \\
\citet{Huijnen2021-nx}, \citeyear{Huijnen2021-nx}                  & x               & x                            &                   &                            &                            &                &                       & x               & x                            & x                 &                            &                 &                &                       &                & x              \\
\citet{smith2021smart}, \citeyear{smith2021smart}                      & x               &                              &                   &                            &                            &                &                       & x               &                              &                   &                            &                 &                &                       &                & x              \\
\citet{catania2021toward}, \citeyear{catania2021toward}                   & x               &                              &                   &                            &                            &                &                       & x               &                              &                   &                            &                 &                &                       &                & x              \\
\hline
\textbf{Count}     & 24              & 12                           & 1                 & 2                          & 1                          & 4              & 5                     & 24              & 15                           & 8                 & 5                          & 5               & 1              & 9                     & 1              & 6         \\
\hline
\end{tabular}
\end{sideways}
\end{table}

\begin{table}[h!]
\centering
\small
\caption{Demographics of the participants involved in the surveyed study including their age, gender, and diagnosis.  If a row has not checks it means that we did not retrieve any information about that specific variable in the corresponding paper.\\
\textit{Legend}: neurodevelopmental disorder (NDD), typically-developed (TD), female (F), male (M), autism spectrum disorder (ASD), intellectual disability (ID), developmental language disorder (DLD).}
\label{tab:demogr}
\begin{sideways}

\begin{tabular}{l|ll|lllll|llll|lll}
\hline
 \multicolumn{1}{l}{{\ul \textbf{Ref}}}     & \multicolumn{2}{c}{{\ul \textbf{Number}}} & \multicolumn{5}{c}{{\ul \textbf{Age}}}                                                                                                    & \multicolumn{4}{c}{{\ul \textbf{Gender}}}                                               & \multicolumn{3}{c}{{\ul \textbf{Diagnosis}}} \\
 \hline
                       & \textbf{NDD}      & \textbf{TD}      & \textbf{Pre} & \textbf{Grade} & \textbf{Teen} & \textbf{Young} & \textbf{Adult} & \textbf{F - NDD} & \textbf{F - TD} & \textbf{M - NDD} & \textbf{M - TD} & \textbf{ASD}  & \textbf{ID}  & \textbf{DLD}  \\
                       & &    & \textbf{schooler} & \textbf{schooler} & & \textbf{ adult} & &  &  &  & &  & &  \\
                       \hline
\citet{mower2011rachel}, \citeyear{mower2011rachel}                  & 2                 &                  &                            & x                              &                       &                              &                      &                       &                      & 2                   &                    & x             &              &               \\
\citet{pop2013social}, \citeyear{pop2013social}                    & 20                &                  & x                          & x                              &                       &                              &                      & 10                    &                      & 10                  &                    & x             &              &               \\
\citet{soleiman2014roboparrot}, \citeyear{soleiman2014roboparrot}                    & 3                 & 2                & x                          &                                &                       &                              &                      &                       &                      &                     &                    & x             &              &               \\
\citet{khosla2015service}, \citeyear{khosla2015service}                       & 2                 &                  &                            &                                &                       &                              & x                    & 1                     &                      & 1                   &                    & x             &              &               \\
\citet{shimaya2016advantages}, \citeyear{shimaya2016advantages}                       & 3                 &                  &                            &                                & x                     &                              &                      & 2                     &                      & 1                   &                    & x             &              &               \\
\citet{bekele2016multimodal}, \citeyear{bekele2016multimodal}                       & 12                &                  &                            &                                & x                     &                              &                      &                       &                      & 12                  &                    & x             &              &               \\
\citet{amanatiadis2017interactive}, \citeyear{amanatiadis2017interactive}                     & 2                 &                  &                            & x                              &                       &                              &                      &                       &                      &                     &                    & x             &              &               \\
\citet{tanaka2017embodied}, \citeyear{tanaka2017embodied}                     & 10                & 18               &                            & x                              & x                     & x                            &                      &                       & 3                    & 10                  & 15                 & x             &              &               \\
\citet{desideri2018using}, \citeyear{desideri2018using}                           & 2                 &                  &                            & x                              &                       &                              &                      &                       &                      & 2                   &                    & x             & x            &               \\
\citet{axelsson2019participatory}, \citeyear{axelsson2019participatory}                          & 10                &                  &                            &                                &                       &                              &                      &                       &                      &                     &                    & x             &              &               \\
\citet{Cao2019-yg}, \citeyear{Cao2019-yg}                       & 15                & 15               & x                          &                                &                       &                              &                      & 2                     & 3                    & 13                  & 12                 & x             &              &               \\
\citet{So2019-vm}, \citeyear{So2019-vm}                    & 23                &                  &                            & x                              &                       &                              &                      & 3                     &                      & 20                  &                    & x             &              &               \\
\citet{Shimaya2019-xg}, \citeyear{Shimaya2019-xg}                      & 2                 & 32               &                            &                                & x                     & x                            & x                    & 2                     & 16                   & 16                  &                    & x             &              &               \\
\citet{Zhang2019-ma}, \citeyear{Zhang2019-ma}                     & 20                & 20               &                            & x                              &                       &                              &                      & 2                     & 3                    & 18                  & 17                 & x             &              &               \\
\citet{Cervera2019-un}, \citeyear{Cervera2019-un}                        & 14                &                  & x                          &                                &                       &                              &                      &                       &                      &                     &                    & x             &              &               \\
\citet{So2020-oq}, \citeyear{So2020-oq}                        & 23                &                  & x                          &                                &                       &                              &                      & 3                     &                      & 20                  &                    & x             &              &               \\
\citet{fachantidis2020tauhe}, \citeyear{fachantidis2020tauhe}                        & 4                 &                  &                            & x                              &                       &                              &                      &                       &                      & 4                   &                    & x             &              &               \\
\citet{marino2020outcomes}, \citeyear{marino2020outcomes}                       & 14                &                  & x                          & x                              &                       &                              &                      & 2                     &                      & 12                  &                    & x             &              &               \\
\citet{Chung2021-fl}, \citeyear{Chung2021-fl}                        & 15                &                  &                            & x                              &                       &                              &                      & 2                     &                      & 13                  &                    & x             &              &               \\
\citet{Van_Otterdijk2020-ww}, \citeyear{Van_Otterdijk2020-ww}                       & 6                 &                  & x                          & x                              &                       &                              &                      & 1                     &                      & 5                   &                    & x             &              &               \\
\citet{spitale2020whom}, \citeyear{spitale2020whom}                         & 3                 & 14               &                            & x                              &                       &                              &                      &                       & 8                    & 3                   & 6                  &               &              & x             \\
\citet{Huijnen2021-nx}, \citeyear{Huijnen2021-nx}                      & 9                 &                  &                            & x                              &                       &                              &                      & 1                     &                      & 8                   &                    & x             &              &               \\
\citet{smith2021smart}, \citeyear{smith2021smart}                          & 43                &                  &                            &                                &                       &                              & x                    &                       &                      &                     &                    &               & x            &               \\
\citet{catania2021toward}, \citeyear{catania2021toward}                       & 9                 &                  & x                          & x                              & x                     &                              &                      & 2                     &                      & 7                   &                    & x             &              &               \\
\hline
\textbf{Count}         & 24                & 6                & 8                          & 14                             & 5                     & 2                            & 3                    & 13                    & 5                    & 19                  & 4                  & 22            & 2            & 1             \\
\textbf{Average}       & 11.08             & 16.83            &                            &                                &                       &                              &                      & 2.54                  & 6.60                 & 9.32                & 12.50              &               &              &               \\
\textbf{St. Deviation} & 9.59              & 8.88             &                            &                                &                       &                              &                      & 2.24                  & 5.08                 & 6.20                & 4.15               &               &              &      \\
\hline
\end{tabular}
\end{sideways}
\end{table}

\begin{table}[h!]
\centering
\small
\caption{Agent's goals in terms of assessing and training and the corresponding skills addressed, such as communication and social, emotional, joint attention, touch confidence, and sensorial stimulation.  If a row has not checks it means that we did not retrieve any information about that specific variable in the corresponding paper.}
\label{tab:skills}
\begin{sideways}

\begin{tabular}{l|ll|lllll}
\hline
\multicolumn{1}{l}{{\ul \textbf{Ref}}} & \multicolumn{2}{c}{{\ul \textbf{Agent's goal}}} & \multicolumn{5}{c}{{\ul \textbf{Skills addressed}}}                                                                                         \\
\hline
                   & \textbf{Assessing}  & \textbf{Training} & \textbf{Communication} & \textbf{Emotional} & \textbf{Joint} & \textbf{Touch } & \textbf{Sensorial } \\
                   & \textbf{}  & \textbf{} & \textbf{and social} & \textbf{control} & \textbf{attention} & \textbf{ confidence} & \textbf{ stimulation} \\
                   \hline
\citet{mower2011rachel}, \citeyear{mower2011rachel}              &                     & x                 & x                                 & x                &                          &                           &                                \\
\citet{pop2013social}, \citeyear{pop2013social}                &                     & x                 & x                                 &                  &                          &                           &                                \\
\citet{soleiman2014roboparrot}, \citeyear{soleiman2014roboparrot}                &                     & x                 & x                                 &                  &                          &                           &                                \\
\citet{khosla2015service}, \citeyear{khosla2015service}                   & x                   &                   & x                                 &                  &                          &                           &                                \\
\citet{shimaya2016advantages}, \citeyear{shimaya2016advantages}                   &                     & x                 & x                                 &                  &                          &                           &                                \\
\citet{bekele2016multimodal}, \citeyear{bekele2016multimodal}                   &                     & x                 & x                                 & x                &                          &                           &                                \\
\citet{amanatiadis2017interactive}, \citeyear{amanatiadis2017interactive}                 &                     & x                 & x                                 &                  & x                        &                           &                                \\
\citet{tanaka2017embodied}, \citeyear{tanaka2017embodied}                 &                     & x                 & x                                 &                  &                          &                           &                                \\
\citet{desideri2018using}, \citeyear{desideri2018using}                       &                     & x                 & x                                 &                  &                          &                           &                                \\
\citet{axelsson2019participatory}, \citeyear{axelsson2019participatory}                      &                     & x                 & x                                 &                  & x                        &                           &                                \\
\citet{Cao2019-yg}, \citeyear{Cao2019-yg}                   & x                   &                   & x                                 &                  & x                        &                           &                                \\
\citet{So2019-vm}, \citeyear{So2019-vm}                &                     & x                 & x                                 &                  &                          &                           &                                \\
\citet{Shimaya2019-xg}, \citeyear{Shimaya2019-xg}                  &                     & x                 & x                                 &                  &                          &                           &                                \\
\citet{Zhang2019-ma}, \citeyear{Zhang2019-ma}                 &                     & x                 & x                                 &                  &                          &                           &                                \\
\citet{Cervera2019-un}, \citeyear{Cervera2019-un}                    &                     & x                 & x                                 &                  &                          &                           &                                \\
\citet{So2020-oq}, \citeyear{So2020-oq}                    &                     & x                 & x                                 &                  & x                        &                           &                                \\
\citet{fachantidis2020tauhe}, \citeyear{fachantidis2020tauhe}                    & x                   &                   & x                                 & x                &                          &                           &                                \\
\citet{marino2020outcomes}, \citeyear{marino2020outcomes}                   &                     & x                 &                                   & x                &                          &                           &                                \\
\citet{Chung2021-fl}, \citeyear{Chung2021-fl}                    &                     & x                 & x                                 &                  &                          &                           &                                \\
\citet{Van_Otterdijk2020-ww}, \citeyear{Van_Otterdijk2020-ww}                   &                     & x                 & x                                 &                  & x                        &                           &                                \\
\citet{spitale2020whom}, \citeyear{spitale2020whom}                     & x                   & x                 & x                                 &                  &                          &                           &                                \\
\citet{Huijnen2021-nx}, \citeyear{Huijnen2021-nx}                  & x                   &                   & x                                 &                  &                          & x                         &                                \\
\citet{smith2021smart}, \citeyear{smith2021smart}                      &                     & x                 & x                                 &                  &                          &                           &                                \\
\citet{catania2021toward}, \citeyear{catania2021toward}                   & x                   &                   & x                                 &                  &                          &                           & x                              \\
\hline

\textbf{Count}     & 6                   & 19                & 23                                & 4                & 5                        & 1                         & 1           \\     
\hline
\end{tabular}
\end{sideways}
\end{table}

\begin{table}[h!]
\centering
\small
\caption{Study designs - such as random control trail, within-subject, between-subject studies - and duration of the empirical study reported in the surveyed papers.  If a row has not checks it means that we did not retrieve any information about that specific variable in the corresponding paper.\\
\textit{Legend}: typically-developed (TD), neurodevelopmental disorder (NDD), random control trail (RCT).}
\label{tab:study}
\begin{sideways}

\begin{tabular}{l|llllllll|ll}
\hline
\multicolumn{1}{l}{{\ul \textbf{Ref}} }    & \multicolumn{8}{c}{{\ul \textbf{Study Design}}}                                                                                                                                    & \multicolumn{2}{c}{{\ul \textbf{Study Duration}}}     \\
\hline
                       & \textbf{TD vs.} & \textbf{RCT} & \textbf{Within-} & \textbf{Between-} & \textbf{Single-} & \textbf{Longitudinal} & \textbf{Exploratory} & \textbf{Pre- } & \textbf{Number} & \textbf{Duration} \\
                        & \textbf{ NDD} & \textbf{} & \textbf{subject} & \textbf{subject} & \textbf{case} & \textbf{} & \textbf{} & \textbf{ Post-} & \textbf{ sessions} & \textbf{(day)} \\
                       \hline
\citet{mower2011rachel}, \citeyear{mower2011rachel}                  &                     &              &                         &                          &                      &                       & x                    &                     & 4                           &                         \\
\citet{pop2013social}, \citeyear{pop2013social}                    &                     & x            &                         & x                        &                      &                       &                      &                     & 6                           & 6                       \\
\citet{soleiman2014roboparrot}, \citeyear{soleiman2014roboparrot}                    & x                   &              &                         &                          & x                    &                       & x                    &                     & 1                           & 1                       \\
\citet{khosla2015service}, \citeyear{khosla2015service}                       &                     &              &                         &                          & x                    & x                     &                      &                     &                             & 270                     \\
\citet{shimaya2016advantages}, \citeyear{shimaya2016advantages}                       &                     &              &                         &                          & x                    &                       &                      & x                   & 4                           & 4                       \\
\citet{bekele2016multimodal}, \citeyear{bekele2016multimodal}                       &                     & x            &                         & x                        &                      &                       &                      & x                   & 5                           &                         \\
\citet{amanatiadis2017interactive}, \citeyear{amanatiadis2017interactive}                     &                     &              &                         &                          & x                    &                       &                      &                     & 1                           & 1                       \\
\citet{tanaka2017embodied}, \citeyear{tanaka2017embodied}                     & x                   &              &                         &                          &                      &                       & x                    & x                   & 1                           & 1                       \\
\citet{desideri2018using}, \citeyear{desideri2018using}                           &                     &              &                         &                          & x                    &                       &                      & x                   & 12                          & 28                      \\
\citet{axelsson2019participatory}, \citeyear{axelsson2019participatory}                          &                     &              & x                       &                          &                      &                       &                      &                     & 1                           & 1                       \\
\citet{Cao2019-yg}, \citeyear{Cao2019-yg}                       & x                   & x            & x                       &                          &                      &                       &                      &                     & 1                           & 1                       \\
\citet{So2019-vm}, \citeyear{So2019-vm}                    &                     & x            &                         & x                        &                      &                       &                      & x                   & 9                           & 63                      \\
\citet{Shimaya2019-xg}, \citeyear{Shimaya2019-xg}                      &                     &              &                         &                          & x                    &                       &                      &                     & 4                           & 4                       \\
\citet{Zhang2019-ma}, \citeyear{Zhang2019-ma}                     & x                   &              &                         &                          &                      &                       & x                    &                     & 1                           & 1                       \\
\citet{Cervera2019-un}, \citeyear{Cervera2019-un}                        &                     & x            &                         & x                        &                      & x                     &                      & x                   & 40                          & 270                     \\
\citet{So2020-oq}, \citeyear{So2020-oq}                        &                     & x            &                         & x                        &                      & x                     &                      & x                   & 9                           & 63                      \\
\citet{fachantidis2020tauhe}, \citeyear{fachantidis2020tauhe}                        &                     &              & x                       &                          & x                    &                       &                      &                     & 8                           & 150                     \\
\citet{marino2020outcomes}, \citeyear{marino2020outcomes}                       &                     & x            &                         & x                        &                      &                       &                      & x                   & 6                           & 84                      \\
\citet{Chung2021-fl}, \citeyear{Chung2021-fl}                        &                     &              &                         &                          & x                    & x                     &                      & x                   & 12                          & 84                      \\
\citet{Van_Otterdijk2020-ww}, \citeyear{Van_Otterdijk2020-ww}                       &                     & x            &                         &                          &                      & x                     &                      &                     & 20                          & 140                     \\
\citet{spitale2020whom}, \citeyear{spitale2020whom}                         & x                   & x            & x                       &                          & x                    &                       &                      &                     & 3                           & 21                      \\
\citet{Huijnen2021-nx}, \citeyear{Huijnen2021-nx}                      &                     & x            & x                       &                          &                      &                       &                      &                     & 4                           & 28                      \\
\citet{smith2021smart}, \citeyear{smith2021smart}                          &                     & x            &                         & x                        &                      & x                     &                      & x                   & 12                          & 84                      \\
\citet{catania2021toward}, \citeyear{catania2021toward}                       &                     &              &                         &                          & x                    &                       & x                    &                     & 2                           & 21                      \\
\hline
\textbf{Count}         & 5                   & 11           & 5                       & 7                        & 10                   & 6                     & 5                    & 10                  & 23                          & 22                      \\
\textbf{Average}       &                     &              &                         &                          &                      &                       &                      &                     & 7.22                        & 60.27                   \\
\textbf{St. Deviation} &                     &              &                         &                          &                      &                       &                      &                     & 8.44                        & 79.66 \\
\hline
\end{tabular}
\end{sideways}
\end{table}

\begin{table}[h!]
\centering
\small
\caption{Study intervention type, defined as the type of interaction for the therapeutic treatment (i.e., task-based or free interaction) and the data collection methodology adopted by the authors in the surveyed papers.  If a row has not checks it means that we did not retrieve any information about that specific variable in the corresponding paper.}
\label{tab:data}
\begin{sideways}

\begin{tabular}{l|ll|llllll}
\hline
\multicolumn{1}{l}{{\ul \textbf{Ref}}}     & \multicolumn{2}{c}{{\ul \textbf{Intervention type}}} & \multicolumn{6}{c}{{\ul \textbf{Data collection }}}                                                                                                                                                   \\
\hline
                       & \textbf{Task-based}      & \textbf{Free}      & \textbf{Interview} & \textbf{Questionnaire} & \textbf{Real-time } & \textbf{In batch } & \textbf{Real-time} & \textbf{In batch} \\
                       & \textbf{}      & \textbf{ interaction}      & \textbf{} & \textbf{} & \textbf{ automated} & \textbf{automated} & \textbf{ manual} & \textbf{manual} \\
                       & \textbf{}      & \textbf{}      & \textbf{} & \textbf{} & \textbf{ data logging} & \textbf{data logging} & \textbf{ data logging} & \textbf{data logging} \\
                       \hline
\citet{mower2011rachel}, \citeyear{mower2011rachel}                  & x  &                                & x                  &                        & x                                         &                                          &                                        & x                                     \\
\citet{pop2013social}, \citeyear{pop2013social}                    & x                        &                                &                    & x                      &                                           &                                          &                                        & x                                     \\
\citet{soleiman2014roboparrot}, \citeyear{soleiman2014roboparrot}                    &                          & x                              &                    &                        &                                           &                                          &                                        & x                                     \\
\citet{khosla2015service}, \citeyear{khosla2015service}                       &                          & x                              &                    &                        & x                                         &                                          &                                        & x                                     \\
\citet{shimaya2016advantages}, \citeyear{shimaya2016advantages}                       &                          & x                              &                    &                        &                                           & x                                        &                                        &                                       \\
\citet{bekele2016multimodal}, \citeyear{bekele2016multimodal}                       & x                        &                                &                    &                        & x                                         &                                          &                                        &                                       \\
\citet{amanatiadis2017interactive}, \citeyear{amanatiadis2017interactive}                     & x                        &                                &                    & x                      &                                           &                                          & x                                      &                                       \\
\citet{tanaka2017embodied}, \citeyear{tanaka2017embodied}                     & x                        &                                &                    &                        & x                                         &                                          &                                        &                                       \\
\citet{desideri2018using}, \citeyear{desideri2018using}                           & x                        &                                &                    &                        & x                                         &                                          &                                        & x                                     \\
\citet{axelsson2019participatory}, \citeyear{axelsson2019participatory}                          & x                        &                                &                    & x                      &                                           &                                          &                                        & x                                     \\
\citet{Cao2019-yg}, \citeyear{Cao2019-yg}                       & x                        &                                &                    &                        & x                                         &                                          &                                        &                                       \\
\citet{So2019-vm}, \citeyear{So2019-vm}                    & x                        &                                &                    &                        &                                           &                                          &                                        & x                                     \\
\citet{Shimaya2019-xg}, \citeyear{Shimaya2019-xg}                      &                          & x                              &                    & x                      &                                           & x                                        &                                        &                                       \\
\citet{Zhang2019-ma}, \citeyear{Zhang2019-ma}                     & x                        &                                & x                  &                        & x                                         &                                          &                                        &                                       \\
\citet{Cervera2019-un}, \citeyear{Cervera2019-un}                        & x                        &                                &                    &                        &                                           &                                          &                                        & x                                     \\
\citet{So2020-oq}, \citeyear{So2020-oq}                        & x                        &                                &                    & x                      &                                           &                                          &                                        &                                       \\
\citet{fachantidis2020tauhe}, \citeyear{fachantidis2020tauhe}                        & x                        &                                & x                  &                        &                                           &                                          & x                                      &                                       \\
\citet{marino2020outcomes}, \citeyear{marino2020outcomes}                       & x                        &                                &                    & x                      &                                           &                                          &                                        &                                       \\
\citet{Chung2021-fl}, \citeyear{Chung2021-fl}                        & x                        &                                &                    &                        &                                           &                                          &                                        & x                                     \\
\citet{Van_Otterdijk2020-ww}, \citeyear{Van_Otterdijk2020-ww}                       & x                        &                                &                    &                        &                                           &                                          &                                        & x                                     \\
\citet{spitale2020whom}, \citeyear{spitale2020whom}                         & x                        &                                &                    & x                      &                                           &                                          &                                        & x                                     \\
\citet{Huijnen2021-nx}, \citeyear{Huijnen2021-nx}                      &                          & x                              & x                  &                        &                                           &                                          &                                        & x                                     \\
\citet{smith2021smart}, \citeyear{smith2021smart}                          &                          & x                              &                    &                        &                                           &                                          &                                        & x                                     \\
\citet{catania2021toward}, \citeyear{catania2021toward}                       & x                        &                                & x                  & x                      & x                                         &                                          &                                        &                                       \\
\hline
\textbf{Count}         & 18                       & 6                              & 5                  & 8                      & 8                                         & 2                                        & 2                                      & 13                                    \\
\hline
\end{tabular}

\end{sideways}
\end{table}
\end{document}